\documentclass[longauth]{aa}
\usepackage{graphicx}
\usepackage{txfonts}
\usepackage[dvipsnames]{xcolor}
\usepackage{orcidlink}
\usepackage{colortbl}
\usepackage[rightcaption]{sidecap}
\sidecaptionvpos{figure}{c}

\usepackage{hyperref}  
\hypersetup{colorlinks=true,linkcolor=[rgb]{1.,0.2,0.2},citecolor=[rgb]{0.1,0.4,1.},filecolor=[rgb]{0.7,0.2,0.2},urlcolor=[rgb]{0.7,0.2,0.2}}

\usepackage{color}
\definecolor{blue}{rgb}{0., 0., 1}

\def\ergs{\rm{erg}\ \rm{s^{-1}}}

\def\Msun{M$_\odot$}

\def\Hei{He\,{\sc i}}

\def\Oiii{[O\,{\sc iii}]}

\def\Oi{[O\,{\sc i}]}
\def\Nii{[N\,{\sc ii}]}

\def\Ha{H$\alpha$}
\def\Hb{H$\beta$}

\def\Pag{Pa$\gamma$}
\def\Mgii{Mg\,{\sc ii}}

\def\Siii{[S\,{\sc iii}]}

\def\kms{km\,s$^{-1}$}

\def\lsim{\mathrel{\rlap{\lower 3pt \hbox{$\sim$}} \raise 2.0pt \hbox{$<$}}}
\def\gsim{\mathrel{\rlap{\lower 3pt \hbox{$\sim$}} \raise 2.0pt \hbox{$>$}}}

\begin{document}

\authorrunning{Loiacono F. et al.}
\titlerunning{No evolution in the number density of little red dots from cosmic dawn to cosmic noon}
\title{No evolution in the number density of little red dots from cosmic dawn to cosmic noon}
%ANCHE: NO DECLINE IN THE NUMBER DENSITY OF LITTLE RED DOTS FROM COSMIC DAWN TO COSMIC NOON
% NO EVOLUTION IN THE NUMBER DENSITY OF LITTLE RED DOTS FROM COSMIC DAWN TO COSMIC NOON
\author{
Federica Loiacono\inst{1}\thanks{ \email{federica.loiacono1@inaf.it}}$^{\orcidlink{0000-0002-8858-6784}}$, 
Roberto Gilli\inst{1}$^{\orcidlink{0000-0001-8121-6177}}$, 
Marco Mignoli\inst{1}$^{\orcidlink{0000-0002-9087-2835}}$, Marcella Brusa\inst{2,1}$^{\orcidlink{0000-0002-5059-6848}}$, Francesco Calura\inst{1}$^{\orcidlink{0000-0002-6175-0871}}$, Marco Chiaberge\inst{3,4}$^{\orcidlink{0000-0003-1564-3802}}$, Andrea Comastri\inst{1}$^{\orcidlink{0000-0003-3451-9970}}$, Quirino D'Amato\inst{5}$^{\orcidlink{0000-0002-9948-0897}}$, Roberto Decarli\inst{1}$^{\orcidlink{0000-0002-2662-8803}}$, Ivan Delvecchio\inst{1}$^{\orcidlink{0000-0001-8706-2252}}$, Kazushi Iwasawa\inst{6,7}$^{\orcidlink{0000-0002-4923-3281}}$, Ignas Juodžbalis\inst{13,14}$^{\orcidlink{0009-0003-7423-8660}}$, Giorgio Lanzuisi\inst{1}$^{\orcidlink{0000-0001-9094-0984}}$, Roberto Maiolino\inst{13,14,15}$^{\orcidlink{0000-0002-4985-3819}}$, Stefano Marchesi\inst{2,8,1}$^{\orcidlink{0000-0001-5544-0749}}$, Giovanni Mazzolari\inst{17,1}$^{\orcidlink{0009-0005-7383-6655}}$, Colin Norman\inst{4,9}$^{\orcidlink{0000-0002-5222-5717}}$, Alessandro Peca\inst{10,11}$^{\orcidlink{0000-0003-2196-3298}}$, Isabella Prandoni\inst{16}$^{\orcidlink{0000-0001-9680-7092}}$, Matteo Sapori\inst{2}$^{\orcidlink{0009-0008-3970-4765}}$, Matilde Signorini\inst{5,12}$^{\orcidlink{0000-0002-8177-6905}}$, Paolo Tozzi\inst{5}$^{\orcidlink{0000-0003-3096-9966}}$, Eros Vanzella\inst{1}$^{\orcidlink{0000-0002-5057-135X}}$, Cristian Vignali\inst{2,1}$^{\orcidlink{0000-0002-8853-9611}}$, Fabio Vito\inst{1}$^{\orcidlink{0000-0003-0680-9305}}$, Gianni Zamorani\inst{1}$^{\orcidlink{0000-0002-2318-301X}}$, Anita Zanella\inst{1}$^{\orcidlink{0000-0001-8600-7008}}$}
\institute{
INAF -- Osservatorio di Astrofisica e Scienza dello Spazio di Bologna, via Gobetti 93/3, I-40129, Bologna, Italy \and Dipartimento di Fisica e Astronomia, Università di Bologna, Via
Gobetti 93/2, I-40129 Bologna, Italy \and Space Telescope Science Institute for the European Space Agency (ESA), ESA Office, 3700 San Martin Drive, Baltimore, MD 21218,
USA \and The William H. Miller III Department of Physics and Astronomy, Johns Hopkins University, Baltimore, MD 21218, USA \and INAF – Arcetri Astrophysical Observatory, Largo E. Fermi 5, I50125, Florence, Italy \and Institut de Ciències del Cosmos (ICCUB), Universitat de Barcelona (IEEC-UB), Martí i Franquès, 1, 08028 Barcelona, Spain \and ICREA, Pg. Lluís Companys 23, 08010 Barcelona, Spain \and Department of Physics and Astronomy, Clemson University, Kinard Lab of Physics, Clemson, SC 29634-0978, USA \and Space Telescope Science Institute, 3700 San Martin Drive, Baltimore, MD, USA \and Eureka Scientific, 2452 Delmer Street, Suite 100, Oakland, CA 94602-3017, USA \and Department of Physics, Yale University, P.O. Box 208120, New Haven, CT 06520, USA \and Dipartimento di Matematica e Fisica, Univeristà di Roma 3, Via della Vasca Navale, 84, 00146 Roma, Italy \and Kavli Institute for Cosmology, University of Cambridge, Madingley Road, Cambridge CB3 OHA, UK \and Cavendish Laboratory – Astrophysics Group, University of Cambridge, 19 JJ Thomson Avenue, Cambridge CB3 OHE, UK \and Department of Physics and Astronomy, University College London, Gower Street, London WC1E 6BT, UK \and INAF - Istituto di Radioastronomia, Via Gobetti 101, I-40129 Bologna, Italy \and Max-Planck-Institut für extraterrestrische Physik (MPE), Gießenbachstraße 1, 85748 Garching, Germany}

\date{}
\abstract{We present our search for little red dots (LRDs) in the '\textit{J1030 field}', a region of the sky around the $z\sim 6.3$ quasar SDSS J1030+0524, observed by the JWST EIGER program.
Over 154 point-like sources selected in a JWST-based photometric catalog, we find five broad line emitters (with $FWHM \gtrsim 1000\ $\kms) that are red ($F200W - F356W > 0$) and are undetected in the X-rays.
%Based on their rest optical and ultraviolet spectral slopes, three of them are confirmed LRDs, one is compatible with the so-called 'little blue dots', and the last one looks like an intermediate object between the two classes of sources. 
%We do not split our sample into sub-populations, and we use all the X-ray silent, red, and compact broad line emitters to derive the bolometric luminosity function (LF) of LRDs at $z = 2.4$ and $z = 4.5$.
We use these sources to derive the bolometric luminosity function (LF) of LRDs at $z = 2.4$ and $z = 4.5$.
%We derive the bolometric luminosity function (LF) of LRDs at $z = 2.4$ and $z = 4.5$. 
At $z = 2.4$, the space density of LRDs is only a factor of $\sim 2$ lower than that of all pre-JWST active galactic nuclei (AGNs) with bolometric luminosity $L_{\rm bol} \gtrsim 3 \times 10^{44}\ \rm erg\ s^{-1}$. 
At $z = 4.5$, our estimate is consistent with those derived for LRDs based on larger areas of the sky.
A similar behaviour is observed in the black hole mass function.
More importantly, we study the number density of LRDs from cosmic dawn to cosmic noon. We find that there is no significant evolution in the abundance of LRDs with $L_{\rm bol} \gtrsim 3 \times 10^{44}\ \rm erg\ s^{-1}$ at $z > 2$. We speculate that the drop at $z < 4$ seen by other studies is due to their sampling of only the bright-end of the LRDs LF. At cosmic noon, the abundance of LRDs is $n = 3.4^{+5.6}_{-2.4} \times 10^{-5}\ \rm Mpc^{-3}$, which is a factor of $\sim 350$ larger than recent model predictions and is comparable with that of X-ray selected AGNs with similar bolometric luminosity.
Our result may imply that, if LRDs are the early, rapid stages of supermassive black hole growth, as suggested by some models, then the formation of black hole seeds can be efficient down to epochs as recent as cosmic noon. Alternatively, LRDs may simply be a high-accretion phase in already mature black holes.}
%If, as suggested by models, LRDs are the early, rapid stages of growth of supermassive black holes, our result implies that the formation of black hole seeds can be efficient up to epochs as recent as cosmic noon.}    
\keywords{quasars: supermassive black holes --- quasars: emission lines --- galaxies: high-redshift --- galaxies: active}
\maketitle    

\section{Introduction} 
\label{sec:intro}
%Discovery and properties of LRDs
One of the most intriguing results of the James Webb Space Telescope (JWST) has been the discovery of a previously unknown population of extragalactic objects called "little red dots" (LRDs; \citealt{matthee24}). These sources are mostly unresolved in JWST/NIRCam images, have red near-infrared (NIR) colors, and show a 'v-shaped' UV-optical spectral energy distribution (SED), with the inflection around the Balmer break wavelength (3645 \AA).    
%Focus on broad lines, AGNs, and odd AGN properties
When looking at their spectra, most LRDs (up to $\sim 60\%$ according to \citealt{greene24}, or even more, see \citealt{hviding25}) show broad emission lines with full-width at half-maximum ($FWHM$) of $1000-2000$ \kms . This fact suggests that LRDs can be powered by active galactic nuclei (AGNs) with black hole masses $M_{\rm BH} \sim 10^7 - 10^8$ \Msun . 
%Eventualmente posso introdurla qui la più vasta famiglia di BLAGNs scoperti da Webb

However, some of their properties are at odds with those of classical AGNs. For example, a significant fraction (up to $10-20 \%$ in JWST-discovered broad line AGNs; BLAGNs \citealt{lin24}) shows prominent blueshifted/redshifted absorptions in Balmer lines (e.g., \citealt{matthee24, maiolino24, lin24, kocevski25}), which are extremely rare in low-redshift AGNs ($\sim 0.1 \%$ incidence; \citealt{i&m25}). 
%Their mid-infrared colors do not show strong evidence for a dusty torus \citep{perezgonz24}, though their stacked SEDs show a rising near-infrared continuum up to $3\ \mu m$ rest-frame \citep{delvecchio25}.
LRDs seem to have lower hot dust emission, typical of the obscuring torus, than normal AGNs, although most of them do show near-IR and mid-IR excess emission, both individually and in stacks \citep{delvecchio25, brazzini26, ji26, lin26, perezgonz26}.
Compared to local supermassive black holes, the black holes in LRDs tend to be overmassive with respect to their host galaxies (e.g., \citealt{harikane23, ubler23, maiolino24, juodzbalis24b, jones25, juodzbalis25}).
More intriguingly, LRDs do not produce either significant X-ray \citep{ananna24, mazzolari24b, yue24, maiolino25} or radio emissions \citep{mazzolari24}.

% AGN-models that account for the odd properties of LRDs
To explain the observed, and in some way exotic, properties of LRDs, \citet{i&m25} propose a scenario in which the Balmer absorption and the strong Balmer break are both produced by the absorption from extremely dense gas along the line of sight (with hydrogen density $n_{\rm H} \gtrsim 10^9\ \rm cm^{-3}$).
Absorption from dense, dust-poor clouds in the broad line region (BLR) might also explain the X-ray weakness in LRDs, provided that the covering factor is significantly larger than in classical AGNs \citep{maiolino25}.
Another scenario proposed to interpret the X-ray weakness invokes super-Eddington accretion onto the black hole \citep{madau24, pacucci24, king25, madau25}.  
Based on detailed analysis of individual LRDs rest-frame optical spectra, some studies interpret their observational features as a result of a "black hole star" \citep{begelman2008}, i.e, a supermassive black hole surrounded by a turbulent, highly dense, and dust-free atmosphere \citep{degraaff25, naidu25}.     

% Caveat broad lines, alternative scenarios: galaxies, globular clusters

Some of the scenarios described above put into question the presence of highly massive black holes at the center of LRDs.
For example, some studies suggest that the exponential profiles observed in the Broad lines of a fraction of LRDs (less than $50\%$ according to \citealt{scholtz26}) may not be the effect of virial gas motions around the black hole but could be instead produced by resonant Balmer or electron scattering, implying that black hole masses estimated through standard calibrations may be biased high, up to a factor of $\sim 100\times $ \citep{naidu25, rusakov25}. However, other works illustrate that exponential profiles can simply arise from virial motions in a stratified BLR \citep{madau26, scholtz26}, and that black hole masses seem consistent with those derived from the single epoch virial estimators \citep{juodzbalis26}. 
%For example, in the black hole star model, the broad profile of the Balmer lines may not simply arise from virial gas motions around the black hole, but could be instead broadened by resonant Balmer or electron scattering, implying that black hole masses estimated through standard calibrations may be biased high, up to a factor of $\sim 100\times $ \citep{naidu25, rusakov25}.
In addition, rather than being explained by absorption from dense gas around the black hole, \citet{labbe24} suggested that the strong Balmer break may be produced by evolved stellar populations. Other scenarios finally explain the spectral properties of LRDs as the sum of UV radiation from very young stars and optical emission from a short-lived supermassive star (with a mass $M > 10^4$ \Msun). In this framework, LRDs would be globular clusters in formation, rather than accreting supermassive black holes, and the broad line profiles would be produced by winds from the supermassive stars at their center \citep{chisholm26}. These examples show how, three years after their discovery, the nature of LRDs remains hazy and necessitates further investigation.

% Little blue dots, relative abundances, and possible explanation
Besides, LRDs are only a part of a vaster population of X-ray silent, broad-line AGNs discovered by JWST. According to some studies, only a small fraction ($10-30 \%$) of these sources is represented by LRDs \citep{hainline25, brazzini26, taylor24, geris26}. The remaining sources lack the 'v-shaped' SED, have blue optical continuum (optical slope $\beta_{\rm opt} < 0$), and they are referred to as little blue dots (LBDs).
According to recent models, LRDs may represent the dust-reddened version, viewed at high-inclination angles of LBDs \citep{Madau&Maiolino26}. 
However, the fraction of LRDs found by other studies can be as high as $60-80\%$ \citep{hviding25, zhuang25}, suggesting that selection effects may play a major role in determining the relative abundance of sub-populations of JWST-discovered BLAGNs. 

% Demography and knowledge gap where to insert this work: forse ci sta meglio dopo Chisholm+26
Regarding their demographics, several studies attempted to estimate the abundance of LRDs and compare it with that of other AGN populations. 
At $z \sim 5-7$, LRDs are $1\ \rm dex$ more abundant than X-ray and UV selected AGNs with similar UV absolute magnitude $M_{\rm UV}$ \citep{kocevski25}. A similar result was found by \citet{kokorev24}, who reported an abundance of LRDs a factor of $\sim 100\times$ larger than that of UV-selected quasars at $6.5 < z < 8.5$ and with $-22 < M_{UV} < -17$.
Concerning the number density of LRDs as a function of time, \citet{kocevski25} find that their redshift distribution sharply declines at $z < 4.5$. The drop in the number density of LRDs at those redshifts was confirmed by \citet{ma25} using ground-based photometric data. On the other hand, \citet{loiacono25}, based on a small sample of spectroscopically confirmed LRDs, find that their abundance at cosmic noon is only a factor of $\sim 2-3$ lower than that of UV-selected quasars with comparable bolometric luminosities, meaning that this new population is still relevant at $z \sim 2-3$. Similarly, \citet{bisigello25} find no drop in the abundance of LRDs between $z \sim 4$ and $z \sim 1.5$, and a decrease only at lower redshifts. This reconciles with recent studies that find a very low number density of LRDs in the local universe $n \sim (1.6 - 5) \times 10^{-9}\ \rm Mpc^{-3}$ \citep{park26, lin26}.
The contradictory results reported at $z < 4$ mean that the demographics of LRDs needs to be further addressed at cosmic noon, which is a crucial epoch in the history of galaxies and their active nuclei \citep{madau&dick}.

% This work
In this work we derive the bolometric luminosity function of LRDs at $z \sim 2.4$ and $z\sim 4.5$. Our sample consists of five compact, red in NIRCam colors, and X-ray silent BLAGNs in the '\textit{J1030 field}', a region of the sky around the $z\sim 6.3$ quasar SDSS J1030+0524 (hereafter, J1030). This area is covered by several multi-wavelength data, including deep radio \citep{damato22} and X-ray \citep{nanni20} observations.
At optical and near-infrared wavelengths, it was imaged by LBT/LBC, CFHT/WIRCam, and \textit{Spitzer}, with both IRAC \citep{annunziatella18} and MIPS, while in the millimeter domain, the field was covered by AzTEC \citep{zeballos18}.
The central $1' \times 1'$ region was observed with HST/ACS, HST/WCF3, VLT/MUSE, and ALMA \citep{stiavelli05, damato20, mignoli20}. The field is also part of the MUSYC survey \citep{gawiser06}.
Finally, the \textit{J1030 field} is one of the targeted regions by the EIGER (Emission-line galaxies and Intergalactic Gas in the Epoch of Reionization; \citealt{kashino23}) program, which observed it with JWST/NIRCam, both in imaging and slitless spectroscopy.  
We discovered the sources analyzed in this work in this JWST dataset, and we use them to derive the number density of LRDs at cosmic noon.

% Structure
This work is organized as follows. In Sect.~\ref{sec:datared} we detail the data reduction and in Sect.~\ref{sec:sel} present the sample selection. We analyze the spectra in Sect.~\ref{sec:analysis} and present the main results in Sect.~\ref{sec:res}. Finally, we discuss our findings in Sect.~\ref{sec:disc} and summarize this work in Sect.~\ref{sec:concl}.

% Cosmology and errors
In this paper we adopt a $\Lambda$CDM cosmology with $\Omega_{\Lambda} = 0.7$, $\Omega_{\rm M} = 0.3$ and an Hubble constant $H_0 = 70\ \rm km\ s^{-1}\ Mpc^{-1}$. The reported errors correspond to the statistical uncertainty of 68\% (1$\sigma$), if not specified differently. 

\section{Data reduction}
\label{sec:datared}
\subsection{The J1030 field seen by EIGER}
\label{sub:dataredgen}
We used the archival data of the 1243 GTO program EIGER (PI: S. J. Lilly). Briefly, this program targets the fields of six $5.9 < z < 7$ quasars (J0100+2802, J0148+0600, J1030+0524, J159-02, J1120+0641, and J1148+5251) with NIRCam imaging (in the $F115W$, $F200W$, and $F356W$ filters) and wide-field slitless spectroscopy (WFSS; $F356W$ filter and $R$ grism); see \citet{kashino23} and \citet{eilers24} for further details. In our analysis, we focused on the \textit{J1030 field}. The images and spectra were acquired in two periods, between May 30 and June 2, 2023, and on May 25, 2024. The field was observed in five telescope visits (visit-1 to visit-5), covering four sky regions that include the quasar, and which are combined into a mosaic.  
\begin{figure*}
\begin{center}
\includegraphics[width=1.\textwidth]{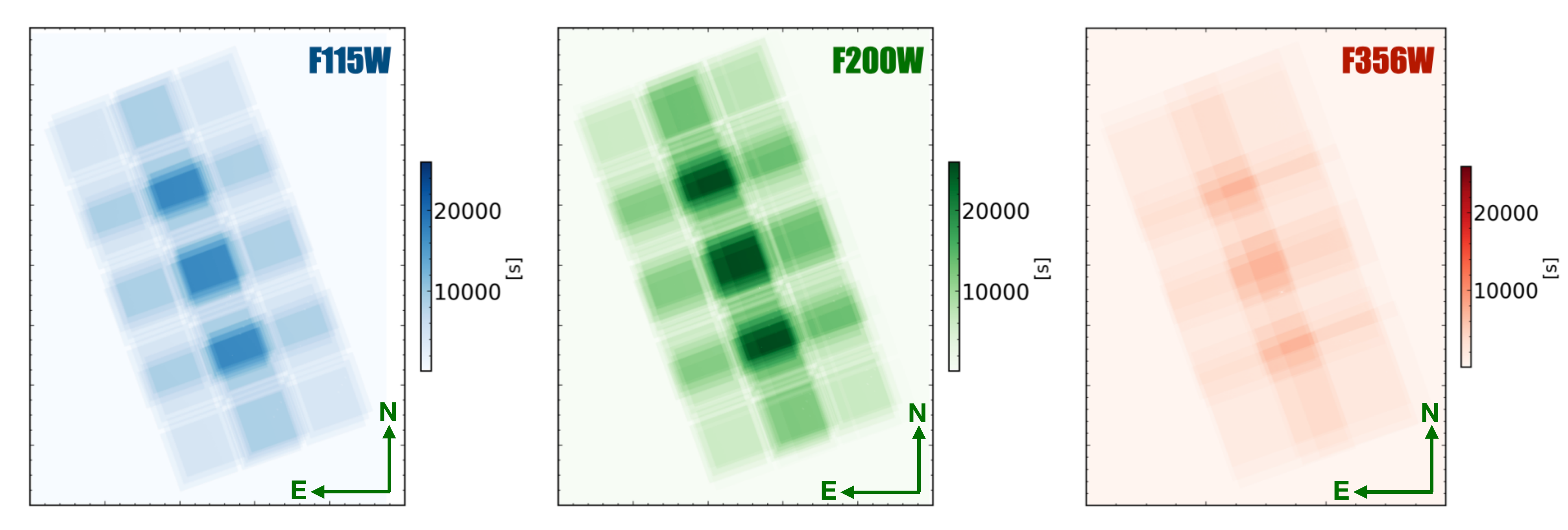}\\
\end{center}
\caption{Exposure maps of the \textit{F115W}, \textit{F200W}, \textit{F356W} imaging mosaics (size $\sim 6.8' \times 3.6'$) of the \textit{J1030 field}. The central tiles reach the maximum exposure time, which is $17500\ \rm s$ (\textit{F115W}), $25800\ \rm s$ (\textit{F200W}), and $7300\ \rm s$ (\textit{F356W}).}
\label{fig:expmap}
\end{figure*}

% General info: images
The NIRCam images form a mosaic of $\sim 6.8' \times 3.6'$ around the quasar. The exposure time varies across the field and reaches in the central part the value of $\sim$17.5~ks, $\sim$25.8~ks, and\ $\sim$7.3\ ks for the $F115W$, $F200W$, and $F356W$ filters, respectively (Figure~\ref{fig:expmap}).

% General info: spectra
The WFSS data of the \textit{J1030 field} cover an area $\sim 17 \ \rm arcmin^2$ at $3.1\ \mu m$ and $\sim 25 \ \rm arcmin^2$ at $3.95\ \mu m$ (see \citealt{kashino23}). The observed spectral range is $3.1-4.0\ \mu m$, and the resolving power $R = \lambda/ \Delta \lambda$ is $\sim 1500$ at the central wavelength. 
The total obseerving time for the WFSS data is $\sim 36.9\rm \ ks$, which also corresponds to the maximum exposure reached in the central region of the field.

\subsection{$F115W$, $F200W$, and $F356W$ mosaics}
\label{sub:dataredimg}
 %%%Reduction
We reduced the imaging data with the 1.14.0 version (\texttt{jwst\_1293.pmap}) of the JWST pipeline \citep{bushouse}. The reduction consists of three stages. We processed the raw data (\textit{"uncal"} files) through the \textit{Stage1} and \textit{Stage2} of the pipeline. 
%1/f noise subtraction
At the end of \textit{Stage2}, we subtracted the correlated readout noise ($1/f$ noise) from the calibrated exposures (\textit{"cal"} files) using the \texttt{Image1overf} code\footnote{\url{https://github.com/chriswillott/jwst}}, developed by Chris Willott. We subtracted the noise per row, per amplifier, and also per column, as the images were also affected by vertical striping.
%Background subtraction
We then computed the 2D background in each exposure, calculating the $\sigma$-clipped median over $200\ \rm px \times 200\ px$ boxes (for the $F115W$ and $F200W$ filters, $400\ \rm px \times 400\ px$ for the $F356W$ one), after masking out the sources in the field. We thus subtracted the 2D background model from each exposure. 

%Wisp subtraction 
Before running \textit{Stage3}, we subtracted the scattered light (i.e., \textit{"wisps"}) from the background-subtracted exposures. This well-known spurious signal affects the A3, A4, B3, and B4 short-wavelength detectors at fixed pixel positions. We subtracted it using the templates made available by STScI\footnote{\url{https://stsci.app.box.com/s/1bymvf1lkrqbdn9rnkluzqk30e8o2bne}}. This procedure effectively removed the wisps from the exposures of all visits but visit-4 for the $F115W$ and $F200W$ filters, and visit-3 (A3 and A4 detectors only) for the $F200W$ filter. The exposures from these visits were further processed by modeling the residual wisp emission with a 2D fit after masking out the signal from sources in the field. The 2D model was hence subtracted from each exposure individually. This procedure efficiently removed wisps from the final $F115W$ and $F200W$ mosaics except for faint residuals in limited areas.   
In addition, we also found wisps in the long wavelength detectors (A5 and B5) for visit-3 and visit-4. STScI wisp templates are not available for these detectors. To remove them, we computed the median over the A5 and B5 exposures (separately) of all visits and subtracted it from each exposure of visit-3 and visit-4.   
Subtracting the median effectively removed wisps from all exposures of visit-3. For visit-4 it was necessary to make a further subtraction, considering the median over the exposures of visit-3 and visit-4 only after masking out the sources in the field. This removed the wisps from the final $F356W$ mosaic effectively, except for limited areas where low residuals are present.

%Astrometry
We then processed the corrected exposures in the \textit{Stage3} of the pipeline. First of all, using the \texttt{Tweakreg} step, we registered the $F356W$ tiles on the multiband $Ks$-selected catalog of J1030 \citep{mazzolari26}, which was previously astrometrized on Gaia stars, and created the final mosaic.
% Pixel size
We selected a pixel scale of $0.03$\arcsec, i.e., half the native pixel size for the $F356W$ mosaic, to match the native pixel scale of the $F115W$ and $F200W$ images, and have pixels with the same size in the three mosaics. 
% Hot pixels
We removed the artifacts due to cosmic rays and detector issues, which we identified by tuning SExtractor \citep{bertin&arnouts96} for detecting hot pixels.
% Offsets
Then, for the $F115W$ and $F200W$ tiles, we created a mosaic for each visit, registering it on the $F356W$ one. We applied an offset in R.A and Dec to each $F115W$ and $F200W$ visit and detector to account for systematic shifts in source positions with the $F356W$ filter.
% Swarp
The final mosaics were obtained with \texttt{Swarp} \citep{bertin2002}, using the "\textit{WHT}" extensions of NIRCam tiles as weight images, to account for the non-homogeneous coverage of the field. 

\begin{figure*}
\begin{center}
\includegraphics[width=0.98\textwidth]{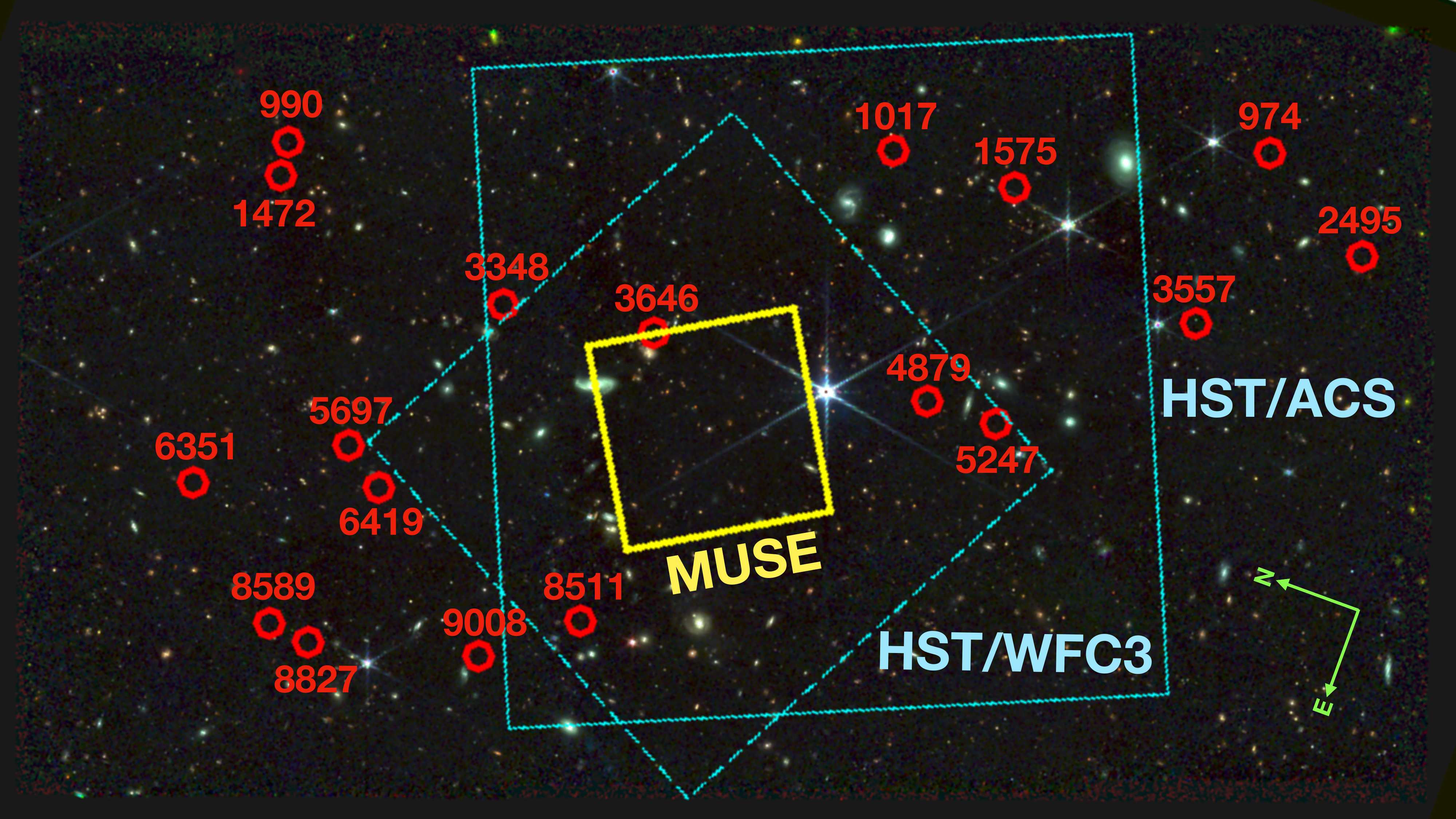}\\
\end{center}
\caption{The JWST/NIRCam $F115W + F200W + F356W$ mosaic of the \textit{J1030 field}. It covers a sky area of $\sim 6.8' \times 3.6'$. We also show the MUSE, HST/ACS, and HST/WFC3 footprints (yellow and cyan, respectively). The red circles show the positions of the 18 point-like sources with color $F200W - F356W > 0$, excluding the X-ray AGNs (see Sect.~\ref{sub:sel} and Fig.~\ref{fig:cc}). These sources are listed in Table~\ref{tab:sample}.}
\label{fig:field}
\end{figure*}

The composite $F115W + F200W + F356W$ mosaic of the \textit{J1030 field} is shown in Fig.~\ref{fig:field}.

\subsection{WFSS data}
\label{sub:dataredwfss}
%Reduction
We processed the raw exposures through the \textit{Stage1} of the pipeline and we applied the World Coordinate System using the \texttt{assign\textunderscore wcs} step in \textit{Stage2}. The products ("\textit{rate}" files) were reduced with the routine\footnote{\url{https://github.com/fengwusun/nircam_grism/}} developed by Fengwu Sun (see \citealt{sun23}). 
%Flat-field
We applied to all exposures the imaging flat-field based on in-flight data, since there is no grism flat-field at the moment.

%Background subtraction
To remove the background, one possibility consists of subtracting a template (for each detector) built as the median over all flat-fielded exposures. However, we could not apply this method, as the data of visit-3 and -4 are contaminated by "ghosts" (i.e., artifacts possibly due to scattered light) in both A5 and B5 modules that strongly affect the median templates. This procedure results indeed in an oversubtraction of the background for the exposures that are not affected by ghosts.
We thus removed the background by subtracting the median value from each column of the flat-fielded exposures, after masking out the spectral traces and the ghosts. We note that, while effective in subtracting the background for all visits, this procedure is not aimed at removing the ghosts, which hence continue to affect some spectral traces in visit-3 and -4.
% 1/f noise subtraction
We did not subtract the $1/f$ noise, which results in stripes along the readout direction, as the readout direction for grism R is the same of the dispersion one.

% Continuum subtraction
For each 2D spectrum, we created a continuum-subtracted version that highlights line emission by applying a median filter along the dispersion direction. %However, we did not use these continuum-subtracted spectra in the following analysis, as they may result in an oversubtraction of the broad lines.

% Astrometry
Finally, we registered the astrometry of the grism exposures on the short-wavelength direct images (i.e., $F115W$ and $F200W$) that were obtained at the same time. This operation relies on the internal alignment between the short and long-wavelength apertures and is necessary to account for small astrometric offsets that, however, may strongly impact the quality of the extracted spectra. 

%Spectral extraction
%We used the coordinates (Tab.~\ref{tab:sample}) in the JWST catalog of J1030 (see Sect~\ref{sub:cat}) to extract the 2D spectra of the BLAGNs studied in this work. All spectra were carefully cleaned of contaminant sources, and they were flux calibrated based on the JWST photometry.
To optimally extract the one-dimensional (1D) spectra of the BLAGNs analyzed in this work, we used the coordinates (Tab.~\ref{tab:sample})  from the JWST-derived photometric catalog. Since it was necessary to preserve the continuum flux of the objects selected for a detailed spectral analysis, the extraction was performed on the original 2D spectra, without applying any median filter along the dispersion axis. If no contaminating sources fell within the 7-pixel (0.44~arcsec) extraction window in the undispersed direction, the 1D spectrum was extracted by simply performing a weighted sum perpendicular to the dispersion axis. 
Conversely, when the spectrum of a contaminating object fell within the extraction window, a dedicated decontamination procedure was performed to secure the highest possible accuracy. We utilized the rows of the contaminant that did not overlap with the target spectrum to precisely model the contaminant's continuum shape. This modeled profile was then scaled and subtracted to yield a clean, 1D  spectrum of the main target. Once the decontaminated spectra of the sample BLAGNs were obtained, we compared the synthetic magnitude derived from the spectrum with the F356W band photometric magnitude from the catalog to perform the spectro-photometric calibration. Thanks to the accuracy of the decontamination process and the choice of the extraction aperture size that fit the point-like nature of the targets, the required photometric correction was always below 15\%.

\section{Sample selection}
\label{sec:sel}
\subsection{JWST catalog of the \textit{J1030 field}}
\label{sub:cat}
We constructed the JWST photometric catalog for the \textit{J1030 field}. To compute the photometry, we used SExtractor \citep{bertin&arnouts96} with the parameter settings DETECT\_MINAREA = 12 and DETECT\_THRESH = 2.5. These relatively conservative detection settings were adopted because the main purpose of the catalog is the extraction of slitless spectra. We performed the analysis in dual-image mode, using the \textit{F356W} band as the detection image, since the WFSS data were obtained in this band. Fluxes in the \textit{F115W} and \textit{F200W} filters were measured in circular apertures of various radii, and using a final reference diameter of 0.18\arcsec. Due to the small differences between the point-spread functions (PSFs) in the three filters ($FWHM=$0.040\arcsec, 0.066\arcsec, 0.116\arcsec from the bluest filter to the reddest one), the $F115W$ and $F200W$ images were not convolved to match the PSF of the detection image. Instead, we applied point-source aperture corrections to evaluate robust photometric colors. We estimated and adopted aperture corrections of 0.60, 0.28, and 0.28 for the \textit{F356W}, \textit{F200W}, and \textit{F115W} filters, respectively. 
%This approach optimizes the use of high signal-to-noise pixels and enhances the ability to detect faint counterparts in the two bluer filters without degrading their PSF quality. 
We used this approach because, while tests performed on both the original and PSF-degraded ancillary images yielded no systematic differences in the photometric colors, utilizing the two bluest filters without degrading their PSF offered several advantages. Specifically, it allowed us to exploit higher signal-to-noise ratio pixels at the location of the detected sources, enhanced the detection capability for faint counterparts, and significantly reduced the number of sources contaminated by neighboring objects.

The final photometric catalog comprises 11,838 sources. Because the J1030 field was not homogeneously covered by the NIRCam observations, we estimated the 3$\sigma$ limiting magnitude per pixel for each available band using the noise maps generated during the reduction process. For the forced photometry performed in the \textit{F200W} and \textit{F115W} images at the positions of objects detected in the \textit{F356W} band, the measured aperture magnitudes were compared with the corresponding detection limits. Whenever a source magnitude was fainter than the corresponding 3$\sigma$ limit, the latter was adopted as an upper-limit value. Following this procedure, 87 objects in \textit{F200W} ($<1\%$) and 506 objects in \textit{F115W} ($4\%$) were classified as flux upper limits among the sources detected in \textit{F356W}.
Due to dithering, the mosaic exhibits non-uniform exposure coverage, leading to spatial variations in the photometric limiting magnitude. Nevertheless, across more than 80\% of the image area, the data reach a 5$\sigma$-limiting magnitude of \textit{F356W}~=~28.5.
Finally, to perform a robust morphological classification of the objects in the catalog, we adopted a two-step process aimed at maximizing the quality of the JWST imaging. The first technique relies on the ratio of an object's total flux to the flux measured within a fixed aperture of D=0.18\arcsec. This method efficiently selects point-like sources down to a flux level at which the distribution of faint, compact galaxies overlaps with that of stellar sources. We performed this analysis in both the detection band (\textit{F356W}) and the ancillary \textit{F200W} band, the latter providing the dual advantage of being deeper for blue objects and offering a higher spatial resolution. We did not extend this analysis to the \textit{F115W} image because the PSF is significantly undersampled in this filter. Lastly, the list of candidate point-like sources selected through this method was analyzed using the IRAF task \textit{psfmeasure}, which estimates the spatial extent of the targets by calculating the enclosed flux with increasing aperture radius. For the final classification, we consistently adopted the band in which each target exhibited the highest signal-to-noise ratio (S/N). This two-step morphological analysis yielded a robust list of 154 objects classified as point-like down to a magnitude of \textit{F356W}$\sim$28. 

\subsection{Color-color and color-magnitude diagrams}
\label{sub:sel}

\begin{figure*}
\begin{center}
\includegraphics[width=0.98\textwidth]{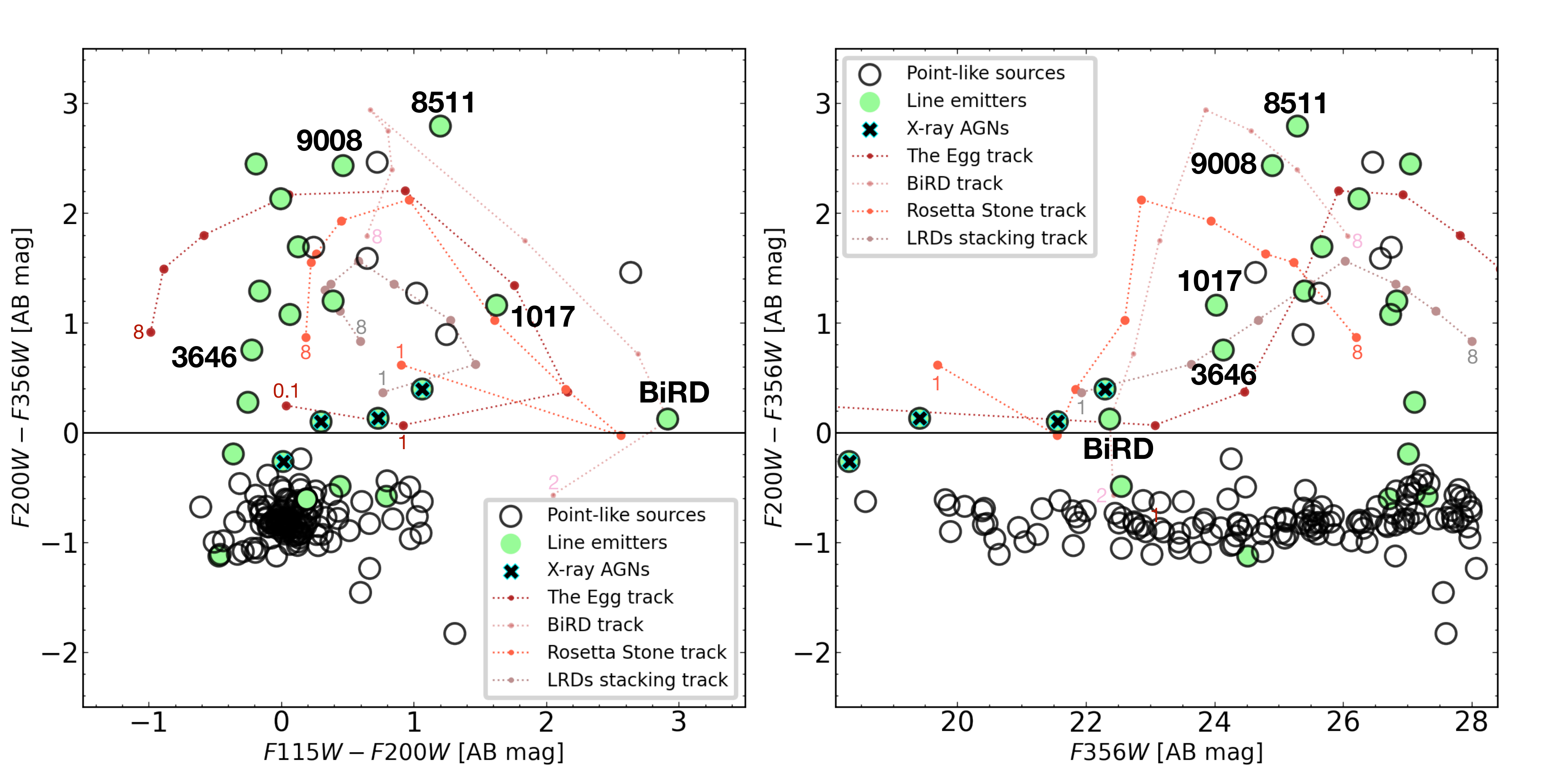}\\
\end{center}
\caption{Color-color (left) and color-magnitude (right) diagrams of point-like sources in the J1030 field. The line emitters are shown with filled green circles. There are four X-ray "classical" AGNs, which are marked with a cross. Red sources are located in the $F200W-F356W > 0$ plane. On the other hand, stars are located at $F200W-F356W < 0$ ("stars locus") of the color-color plot. We show the redshift tracks for some LRDs ("The Egg", \citealt{lin26}; "BiRD", \citealt{loiacono25}; "Rosetta stone", \citealt{juodzbalis24}; stacked sample from \citealt{delvecchio25}) from $z_{\rm min} = 1$ to $z_{\rm max} = 8$ at steps of $\Delta z = 1$, except for BiRD, for which $z_{\rm min} = 2$. For the local LRD, The Egg, we also show the $z = 0.1$ value. These tracks likely identify LRDs at various redshifts in the two diagrams, although some exceptions are present (see Sect.~\ref{sub:sel}). We labeled the broad line, X-ray silent AGNs studied in this work with their names (see Sect.~\ref{sub:blagn}).}
\label{fig:cc}
\end{figure*}

%color-color and color-magnitude diagrams
We plot the $F200W - F356W$ color vs the $F115W - F200W$ color of the 154 point-like sources in the field (Fig.~\ref{fig:cc}, left panel).
We fill in green the symbols corresponding to sources with at least one emission line in the WFSS spectrum extracted at the source position.
We also show the color ($F200W - F356W$) - magnitude ($F356W$) diagram, which highlights the $F356W$ magnitudes of the objects (Fig.~\ref{fig:cc}, right panel).
In Fig.~\ref{fig:cc} (left), the point-like sources are distributed across two distinct regions, characterized by $F200W-F356W$ colors roughly below and above $0$, respectively. The former population corresponds to "blue" sources, which likely represent stars (forming the ``star locus''). To interpret the nature of the second group ("red" sources; 21 in total), we show the color-color track of some LRDs from the literature ("Rosetta stone", \citealt{juodzbalis24}; stacking of LRDs, \citealt{delvecchio25}; "BiRD", \citealt{loiacono25}; "The Egg" or "The Lord", \citealt{lin26, ji26}) computed from their SED as a function of redshift. We can see that the tracks overlap with most of the point-like sources with $F200W - F356W > 0$, suggesting that at least part of them are likely LRDs at various redshifts.

Two of the authors (FL, MM) independently analyzed the WFSS spectra of all the point-like sources shown in Fig.~\ref{fig:cc}, using both the original and decontaminated 1D and 2D spectra. 
% Red sources with lines
For the red sources with more than one line in the grism spectrum, we unambiguously attributed a redshift (see Tab.~\ref{tab:sample}). Based on \Hei\ and \Pag, we found four emitters at $2.3 < z < 2.5$, while, based on the \Oiii\ $\lambda \lambda 4959,5007$ and \Hb\ lines, we identified three emitters at $ 5.7 < z < 6.4$.
For the sources with one line only, when possible, we used the tracks and the inflection in the spectral energy distribution (see Fig.~\ref{fig:sed} in the appendix) to infer the most reasonable nature of the line. In this way, we found three possible \Ha\ emitters at $z \sim 4.4-4.5$.
% Red sources without lines
The nature of the remaining red, point-like sources remains ambiguous. The lack of prominent emission lines, combined with the faintness of the sources themselves and their non-detection in non-JWST photometric bands (both from HST and ground-based facilities), also precludes a search for the potential Balmer break. Consequently, this prevents both the determination of their redshifts and the characterization of their intrinsic nature (e.g., whether or not they harbor an AGN).

% X-ray AGNs
In addition, we found four sources (circles marked with a cross in Fig.~\ref{fig:cc}) that are known X-ray AGNs in the \citet{marchesi21} catalog of $Chandra$ J1030 identifications\footnote{see also \url{http://j1030-field.oas.inaf.it/xray_redshift_J1030.html}}. These objects have $-0.3 < F200W - F356W < 0.5$, and are therefore located between the stellar locus and the cloud of red sources. Also, they are at the bright end of the F356W magnitude distribution (Fig.~\ref{fig:cc}, right). The three X-ray AGNs with bluer colors (XID 10, 21, 30 in \citealt{marchesi21}) are standard, broad line AGNs with spectroscopic redshifts in the range of $z = 1.2-6.3$ (XID 21 is the $z=6.3$ J1030+0524 quasar itself). The fourth X-ray source (i.e., the one with the reddest $F200W - F356W$ color) is a narrow line AGN at $z_{\rm spec}= 2.4$ (XID 31). 

%Drop-out
We also identified an $i$-band dropout ($5.4 < z < 6.6$) just outside the LRD tracks. This source, ID 5247, was discovered by \citet{kim09} through HST/ACS imaging (see Object A8 in their Table), and the narrow line in the WFSS data is likely \Oiii\ $\lambda 5007$ at $z = 5.61$, with the \Oiii\ $\lambda 4959$ and \Hb\ lines possibly too faint to be detected.

% Line emitters in the stars locus
Finally, we found five single-line emitters in the star locus. %\footnote{We found further single-line emitters in the star locus that were likely due to contaminant sources. Therefore, we do not show them in Fig.~\ref{fig:cc}.} 
For these sources, we carefully examined the SED and the spectral trace to exclude the possibility that the line is associated with a contaminant object. Due to the detection of a single line only and the inconclusive photometric information, the nature of these sources remains unclear.

The 18 point-like sources with $F200W - F356W > 0$ (after removing the X-ray AGNs) are listed in Table~\ref{tab:sample}.

\begin{table*}
\caption{The 18 point-like sources with $F200W-F356W > 0$ in the color-color diagram (Fig.~\ref{fig:cc}, left), excluding the X-ray detected AGNs. These sources include confirmed LRDs (BiRD, ID 8511, ID 9008), a confirmed LBD (ID 3646), and an object with intermediate properties (ID 1017), which all have broad lines (IDs in boldface). In addition, there are six line emitters for which there is no evidence for a broad line. The nature of the remaining seven red sources is unclear due to the lack of a spectroscopic identification and a well-characterized SED. The sources are sorted according to their increasing $F356W$ magnitude and are also indicated in Fig.~\ref{fig:field}.}  
\label{tab:sample}      
\centering                                
%\scalebox{0.8}{
\begin{tabular}{c c c c c c c c} 
\hline
ID & R.A. & Dec & Line(s) & $z_{\rm spec}$ &$F115W$&$F200W$&$F356W$ \\[1mm] 
     &(J2000) &(J2000) & & &[AB mag]&[AB mag]&[AB mag] \\
\hline  
\textbf{974 (BiRD)}  & 157.57507 & 5.37915  & \Hei , \Pag , \Oi       & 2.33           & $25.41\pm0.02$ & $22.49\pm0.01$ & $22.37\pm0.03$  \\[1mm]
\textbf{1017} & 157.58558 & 5.40887  & \Hei , \Pag             & 2.50           & $26.82\pm0.03$ & $25.19\pm0.01$ & $24.03\pm0.03$  \\[1mm]
\textbf{3646} & 157.60693 & 5.42235  & \Hei , \Pag , $\rm Pa\delta$, \Siii       & 2.41           & $24.67\pm0.02$ & $24.89\pm0.01$ & $24.14\pm0.03$  \\[1mm]
1575 & 157.58518 & 5.39816  & \textunderscore                    &\textunderscore & $28.73\pm0.06$ & $26.10\pm0.01$ & $24.64\pm0.03$  \\[1mm] 
\textbf{9008} & 157.63721 & 5.42677  & \Ha                    & 4.50            & $27.80\pm0.04$ & $27.33\pm0.02$ & $24.90\pm0.03$  \\[1mm]
\textbf{8511} & 157.63166 & 5.41995  & \Ha                    & 4.50            & $29.3\pm0.2$ & $28.08\pm0.05$ & $25.29\pm0.03$  \\[1mm]
2495 & 157.58051 & 5.36899  & \textunderscore                    &\textunderscore & $27.52\pm0.04$ & $26.27\pm0.01$ & $25.38\pm0.03$  \\[1mm]
8827 & 157.64115 & 5.44069  & \Oiii , \Hb            &6.4             & $26.53\pm0.02$ & $26.69\pm0.02$ & $25.40\pm0.03$  \\[1mm]
990  & 157.60228 & 5.45633  & \textunderscore                    &\textunderscore & $27.92\pm0.04$ & $26.90\pm0.02$ & $25.63\pm0.03$  \\[1mm]
4879 & 157.60447 & 5.39907  & \Oiii, \Hb              & 5.95           & $27.49\pm0.03$ & $27.36\pm0.02$ & $25.67\pm0.03$  \\[1mm]
3348 & 157.60895 & 5.43500  & \Ha                    & 4.39           & $28.37\pm0.06$ & $28.38\pm0.05$ & $26.25\pm0.03$  \\[1mm] 
3557 & 157.59061 & 5.38026  & \textunderscore                    &\textunderscore & $29.6\pm0.4$ & $28.9\pm0.1$ & $26.46\pm0.03$  \\[1mm]
5697 & 157.62438 & 5.44310  & \textunderscore                    &\textunderscore & $28.82\pm0.06$ & $28.17\pm0.03$ & $26.58\pm0.03$  \\[1mm]
6419 & 157.62694 & 5.43941  & \Oiii , \Hb            &5.7             & $27.88\pm0.03$ & $27.81\pm0.02$ & $26.74\pm0.03$  \\[1mm]
1472 & 157.60510 & 5.45610  & \textunderscore                    &\textunderscore & $28.68\pm0.08$ & $28.44\pm0.06$ & $26.75\pm0.03$  \\[1mm]   
6351 & 157.63181 & 5.45420  & \Hei , \Pag                     &2.52               & $28.43\pm0.05$ & $28.03\pm0.04$ & $26.83\pm0.03$  \\[1mm]  
8589 & 157.64073 & 5.44417  & \textunderscore                   &\textunderscore & $29.3\pm0.3$ & $29.5\pm0.2$ & $27.05\pm0.03$  \\[1mm]  
5247$^{(a)}$ & 157.60412 & 5.39300  & \Oiii  &5.61            & $27.14\pm0.02$ & $27.39\pm0.01$ & $27.11\pm0.03$  \\[1mm] 
\hline 
\end{tabular}
%}
\tablefoot{$^{(a)}$ This is object A8 in the sample of HST/ACS $i$-band dropouts in \citet{kim09} (see their Table 3)}
\end{table*}

\subsection{X-ray silent, compact, broad line emitters}
\label{sub:blagn}
% Selection
We select our sample based on these criteria:
\begin{itemize}

\item[(i)] point-like morphology in the $F356W$ and $F200W$ images.\\

\item[(ii)] presence of broad lines (i.e., with a $FWHM > 1000$ \kms).\\

\item[(iii)] lack of X-ray emission.\\
    
\end{itemize}

The resulting sample is made up of five sources that are likely BLAGNs.
% Cosmic noon line emitters and their LRD/LBD nature
We found evidence for a broad \Hei\ $10830$ \AA\ line in three out of four line emitters at $z\sim 2.4$. They are ID 1017, ID 3646, and 'BiRD' (see \citealt{loiacono25} for the latter source).
The fourth emitting source at $z\sim 2.4$ (ID 6351 in Table~\ref{tab:sample}) does not show a significant broad emission. However, its spectrum is affected by a strong contamination, which makes the fit results inconclusive.
%BLAGNs and their nature as LRD or LBD
%We focused on the broad line emitters that are not standard, X-ray AGNs. We found evidence of a broad \Hei\ $10830$ \AA\ line (i.e., with a $FWHM > 1000$ \kms ) in three out of four line emitters at $z\sim 2.4$. We named them 'ID 1017', '\textit{Sandro}', and 'BiRD' (see \citealt{loiacono25} for the latter source).
%The fourth emitting source at $z\sim 2.4$ (ID 6351 in Table~\ref{tab:sample}) does not show a significant broad emission. However, its spectrum is affected by a strong contamination, which makes the fit results inconclusive. 

According to their measured UV and optical slopes ($\beta_{UV}$ and $\beta_{opt}$, where $f_\lambda\propto\lambda^\beta$), and following standard conventions (e.g. \citealt{brazzini26}) BiRD is an LRD, while ID 3646 is an LBD. Based on its mildly V-shaped SED ($\beta_{\rm opt} \sim 0$), ID 1017 is at the border between LRDs and LBDs. The UV and optical spectral slopes were computed by fitting simple power-laws to the available JWST and ground-based photometric datapoints  blue-ward and red-ward of the redshifted Balmer break wavelength, respectively (see \citealt{mazzolari26} for a description of the available multi-band photometry). The measured slopes are reported in Table~\ref{tab:phys} and shown in Fig.~\ref{fig:uvoptslopes}.

%We found evidence of a broad component also in two out of three \Ha\ emitters at $z\sim 4.5$ (ID 8511 and ID 9008).
The last two selected broad line emitters are found in two out of three \Ha\ emitting sources at $z\sim 4.5$ (ID 8511 and ID 9008). 
The third \Ha\ emitter shows a $FWHM \sim 700$ \kms\ which is below our adopted threshold of $1000$ \kms\ (see also \citealt{matthee24}).
We classified both ID 8511 and ID 9008 as LRDs based on their UV and optical slopes (see Sect.~\ref{sub:redvsblue}).

%LRDs without broad lines
The \Oiii + \Hb\ emitters in Table~\ref{tab:sample} do not show evidence for broad components. However, this could be due to the low signal-to-noise ratio ($S/N$) of the \Hb\ emission, and analysis of deeper data would be necessary to assess the nature of these objects.

We report in the following section the analysis of ID 1017 and ID 3646, as BiRD was extensively studied by \citet{loiacono25}.
We also study the two broad \Ha\ emitters at $z\sim 4.5$ (both classified as LRDs based on their UV and optical slopes, see Tab.~\ref{tab:phys}).

\section{Analysis}
\label{sec:analysis}

\subsection{Spectral line fitting of X-ray silent BLAGNs at $z\sim 2.4$}
\label{sub:fit_birds}

%FIT: PROCEDURE BASED ON HEI, SIGNIFICANCE OF HEI BROAD COMPONENT
We fit the \Hei\ emission using both a narrow and a broad Gaussian component. We assessed the significance of the \Hei\ broad component using the Bayesian Information Criterion (BIC), defined as
\begin{equation}
    BIC = \chi^2 + k \ln{N} 
\end{equation}
where $\chi^2$ is the Chi-square, $k$ is the number of fitted parameters, and $N$ is the number of fitted data points. We estimated $\Delta BIC = BIC_N - BIC_{N+B}$, which is the difference between the $BIC$ that we obtained using a narrow component only ($BIC_N$) and both a narrow and a broad component ($BIC_{N+B}$). According to \citet{liddle07}, a $\Delta \rm BIC = 5$ means a 'strong' improvement of the model. The evidence in favor of the model with the lowest $BIC$ becomes 'decisive' if $\Delta \rm BIC = 10$. We obtain $\Delta \rm BIC = 11$ and $\Delta \rm BIC = 5620$ for ID 1017 and ID 3646, respectively, which ensures the reliability of the \Hei\ broad component in both sources.

%PRIORS, REALIZATIONS, ETC., RESULTS AND ERRORS
% INSTRUMENTAL BROADENING
We used the Python package \texttt{emcee} \citep{foreman13} based on a Markov chain Monte Carlo (MCMC) method to perform the line fit. In addition to the \Hei\ line, we fit the continuum emission with a linear function. We used uniform priors for all the fitted parameters, with the $FWHM$ of the narrow and broad components being lower than $1000$ and $10000$ \kms, respectively.
We then fit the \Pag\ emission (and the underlying continuum) separately using a narrow and a broad Gaussian component. In this case, we fixed the $FWHM$ of the two components to the \Hei\ values because of the low $S/N$ of the \Pag\ line. 
%We use this approach because of the low $S/N$ of the \Pag\ line, under the assumption that broad emission affects both helium and hydrogen, tracing gas in the broad line region (BLR).

\begin{figure*}
\begin{center}
\includegraphics[width=0.88\textwidth]{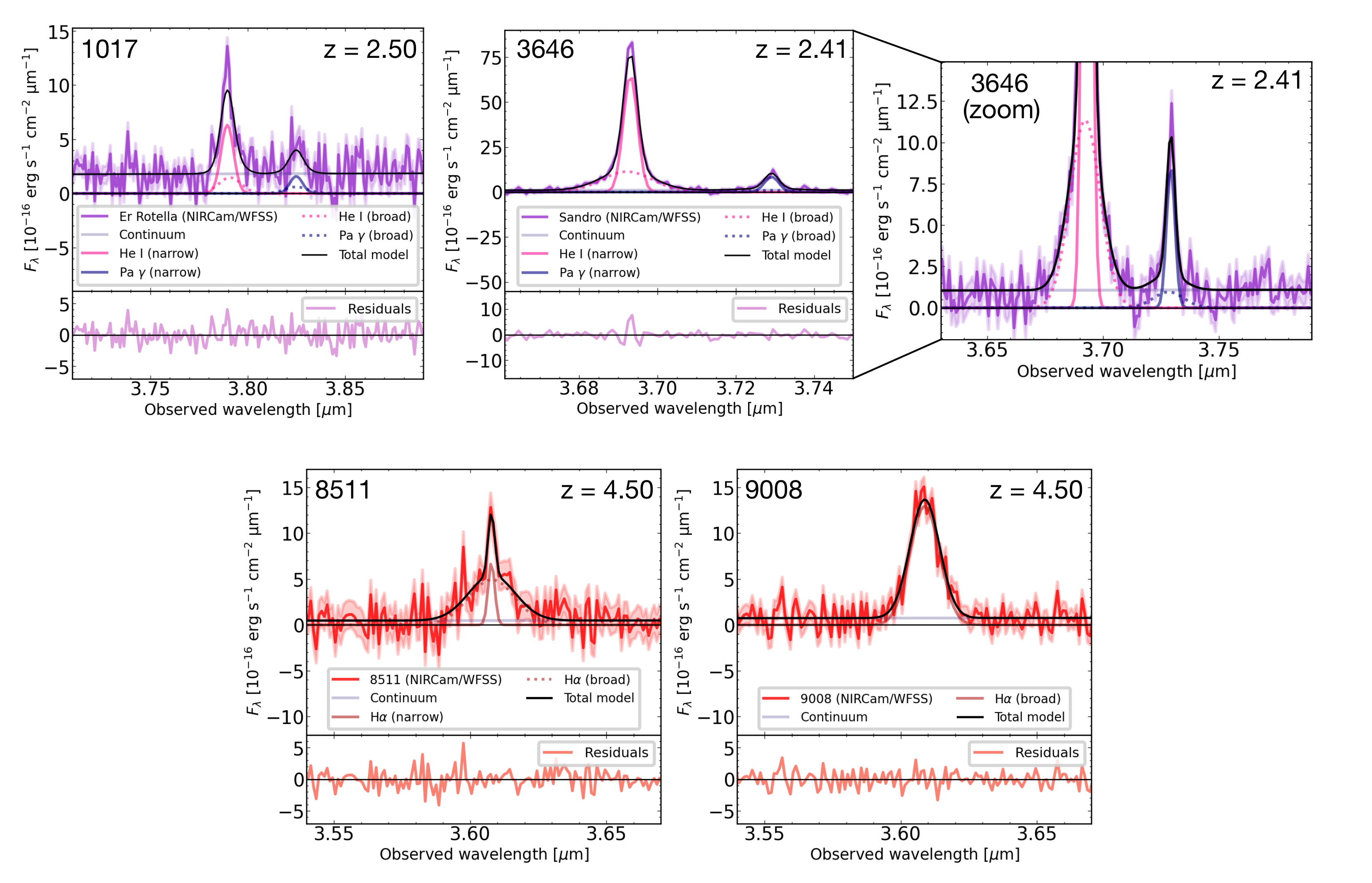}\\
\end{center}
\caption{Line fit of the X-ray silent BLAGNs at $z\sim 2.4$ (ID 1017 and ID 3646) and at $z\sim 4.5$ (ID 8511 and ID 9008). We fit the \Hei\ and \Pag\ emission in ID 1017 and ID 3646 using both a narrow (solid lines) and a broad (dotted lines) Gaussian component. For ID 3646, we also show a zoomed panel to highlight the \Pag\ broad component.
We fit the \Ha\ emission of ID 8511 with both a narrow (solid line) and a broad (dotted line) component. We fit the \Ha\ emission of ID 9008 with a broad component only. We fit a linear continuum to all sources. For each fit, we also show the residuals in the bottom panel.}
\label{fig:fit}
\end{figure*}

We estimated the "best" values as the median of $10000$ realizations and the errors as the 16th and 84th percentiles. Table~\ref{tab:fit_birds} shows the quantities that we derived from the line fit of ID 1017 and ID 3646. The global model is shown in Fig.~\ref{fig:fit}.

We corrected the $FWHM$ of the narrow components for the instrumental broadening, by subtracting in quadrature the instrument $FWHM$, which is for the $R$ grism $FWHM_{\rm instr} \sim 83$ \kms\ at the line wavelengths. For the broad components, this correction is irrelevant.

% COMMENTS ON THE FITTED QUANTITIES 
Based on the \Hei\ narrow component, we estimate the redshift of ID 3646 and ID 1017 to be $z = 2.41004 \pm 0.00008$ and $z = 2.499 \pm 0.001$, respectively. For ID 1017, the velocity shifts $\Delta v$ with respect to the narrow \Hei\ (see Tab.~\ref{tab:fit_birds}) are dominated by large errors for both narrow \Pag\ and broad \Hei\ and \Pag, due to the low $S/N$ of the spectrum. The $FWHM$ of the broad lines is $\sim 1100$ \kms\ and $\sim 1200$ \kms\ for ID 1017 (with a much
larger uncertainty) and ID 3646, respectively.
%as well as the $FWHM$ of the broad components, due to the low $S/N$ of the spectrum.   

%COMMENTS ABOUT FURTHER TESTS: FURTHER NARROW COMPONENT IN SANDRO
We also note that the \Hei\ fitted model of ID 3646 shows a residual near the line center (Fig.~\ref{fig:fit}). This residual can be modeled by adding a further narrow component with $FWHM < 300$ \kms, possibly due to star formation in the host galaxy. However, we do not include this component in the final fit as it does not affect any of the fitted quantities significantly.
%SIGNIFICANCE OF BROAD PASCHEN GAMMA
% Direi di non dire niente: ho già detto che il suo S/N è basso e dunque baso il fit sull'elio
% PERÒ MI SERVE PARLARNE PER L'UPPER LIMIT ---> ne parlo NEL PARAGRAFO APPOSITO  

\begin{table}
\caption{X-ray silent and compact BLAGNs at $z\sim 2.4$: quantities derived from the line fit of ID 1017 and ID 3646}  
\label{tab:fit_birds}      
\centering                                
\scalebox{0.8}{
\begin{tabular}{c c c c}  
\hline
\multicolumn{4}{|c|}{ID 1017} \\
\hline
Line & Flux & $\Delta v$ & FWHM \\[1mm] 
     & ($10^{-18}\ \rm erg\ s^{-1}\ cm^{-2}$) & (\kms) & (\kms) \\
\hline  
\Hei\ (narrow) & $5_{-3}^{+2}$ & $\_$ & $599_{-305}^{+170}$ \\[1mm] 
\Pag\ (narrow) & $1.2_{-0.9}^{+1.2}$ & $-974_{-892}^{+1027}$ & $599(*)$ \\[1mm]
\Hei\ (broad) & $3_{-2}^{+3}$ & $628_{-1710}^{+2189}$ & $1073_{-618}^{+4352}$ \\[1mm] 
\Pag\ (broad) & $1.3_{-1.1}^{+5.3}$ & $-977_{-2242}^{+2998}$ & $1073(*)$ \\[1mm]
\hline
\multicolumn{4}{|c|}{ID 3646} \\
\hline
\Hei\ (narrow) & $29.5_{-1.8}^{+1.9}$ & $\_$ & $336_{-18}^{+20}$ \\[1mm] 
\Pag\ (narrow) & $3.9_{-0.5}^{+0.5}$ & $-498_{-74}^{+65}$ & $336(*)$ \\[1mm]
\Hei\ (broad) & $18.5_{-1.4}^{+1.4}$ & $-245_{-90}^{+85}$ & $1237_{-76}^{+85}$ \\[1mm] 
\Pag\ (broad) & $3.7_{-1.9}^{+2.1}$ & $-763_{-907}^{+878}$ & $1237(*)$ \\[1mm]
\hline 
\end{tabular}}
\tablefoot{The errors are computed from the 16th and 84th percentiles of the fitted parameters in the MCMC fit. The velocity shift $\Delta v$ is estimated with respect to the \Hei\ narrow component. Here $(*)$ means that the quantity was fixed to the \Hei\ value.}
\end{table}

\subsection{Spectral line fitting of the X-ray silent BLAGNs at $z\sim 4.5$}
\label{sub:fit_lrds}

We fit the spectrum of ID 8511 and ID 9008 with a linear continuum and one or two Gaussian components to account for the \Ha\ emission. We used the same priors and number of realizations of the $z\sim 2.4$ objects.
For ID 8511, we found a significant broad component ($\Delta\rm BIC = 9$) with an $FWHM\sim 1700$ \kms. No velocity shift $\Delta v$ is found between the broad and narrow \Ha\ line.
In case of ID 9008, the solution with only one component ($\rm BIC = 167$) is favored over that with two components ($\rm BIC = 177$). The $FWHM$ of the \Ha\ line is  $\sim 1100$ \kms in this case.
We estimated a redshift $z = 4.4972 \pm 0.0005 $ for ID 8511, derived from the narrow \Ha\ component, and a redshift $z = 4.4987 \pm 0.0004$ for ID 9008.
We report the quantities estimated from the fit in Table~\ref{tab:fit_lrds} while we show the fitted model in Fig.~\ref{fig:fit}.

\begin{table}
\caption{X-ray silent and compact BLAGNs at $z\sim 4.5$: quantities derived from the line fit of ID 8511 and ID 9008}  
\label{tab:fit_lrds}      
\centering                               
\scalebox{0.8}{
\begin{tabular}{c c c c}  
\hline
\multicolumn{4}{|c|}{ID 8511} \\
\hline
Line & Flux & $\Delta v$ & FWHM \\[1mm] 
     & ($10^{-18}\ \rm erg\ s^{-1}\ cm^{-2}$) & (\kms) & (\kms) \\
\hline  
\Ha\ (narrow) & $2.1_{-0.6}^{+0.8}$ & $\_$ & $223_{-62}^{+99}$ \\[1mm] 
\Ha\ (broad) & $10_{-1}^{+1}$ & $81_{-522}^{+533}$ & $1668_{-243}^{+335}$ \\[1mm] 
\hline
\multicolumn{4}{|c|}{ID 9008} \\
\hline
\Ha\ (narrow) & $\_$ & $\_$ & $\_$ \\[1mm] 
\Ha\ (broad) & $17.7_{-0.8}^{+0.9}$ & $\_$ & $1067_{-52}^{+59}$ \\[1mm] 
\hline 
\end{tabular}}
\tablefoot{The errors are computed from the 16th and 84th percentiles of the fitted parameters in the MCMC fit. The velocity shift $\Delta v$ is estimated with respect to the \Ha\ narrow component.}
\end{table}

\section{Results}
\label{sec:res}

\subsection{Bolometric luminosity, X-ray and radio properties}
\label{sub:xrp}

%BOLOMETRIC LUMINOSITY
We derive the bolometric luminosity from the broad \Ha\ emission of the sources. For ID 1017 and ID 3646, i.e., the sources where this line is not covered by our data, we inferred the \Ha\ luminosity from the measured broad \Pag\ line assuming Case B recombination. We note that Case B may not apply in the high-density regime of BLR \citep{ilic12}. However, we verified that this scenario is valid for the Rosetta Stone, a $z\sim 2.26$ LRD where both \Pag\ and \Ha\ lines were observed (\citealt{juodzbalis24}; see \citealt{loiacono25} for more details).
Assuming a flux ratio $\sim 31.7$ between \Ha\ and \Pag\ \citep{osterbrock89}, we estimate an intrinsic \Ha\ luminosity of $L(\rm H\alpha) = 2^{+3}_{-1.7} \times 10^{42}\ erg\ s^{-1}$ and $L(\rm H\alpha) = 5^{+3}_{-3} \times 10^{42}\ erg\ s^{-1}$ for ID 1017 and ID 3646, respectively.
On the other hand, the \Ha\ luminosity was directly derived from the observed broad line for the $z\sim 4.5$ sources. We find $L(\rm H\alpha) = 2.1^{+0.3}_{-0.3} \times 10^{42}\ erg\ s^{-1}$ (ID 8511) and $L(\rm H\alpha) = 3.6^{+0.2}_{-0.2} \times 10^{42}\ erg\ s^{-1}$ (ID 9008). The line luminosities, together with other physical quantities, are summarized in Table~\ref{tab:phys}.

%CAVEAT: EXTINCTION CORRECTION
We note that while the inferred \Ha\ luminosity for ID 1017 and ID 3646 takes into account dust extinction, with the caveat that it assumes the same attenuation of the Rosetta Stone (see \citealt{loiacono25} for further details), the observed \Ha\ in the $z\sim 4.5$ sources is not extinction corrected. This implies a possible underestimate of both bolometric luminosity and black hole mass of ID 8511 and ID 9008. We discuss the implications of dust extinction correction to the number density of LRDs at $z \sim 4.5$ in Sect.~\ref{sub:spacedendisc}.

We then converted the \Ha\ luminosity to the bolometric one using the relation of \citet{stern12}

\begin{equation}
L_{\rm bol} = 130^{\times2.4}_{\div 2.4} \; L_{\rm H\alpha,broad}    
\label{eq:stern}
\end{equation}

The bolometric luminosities span the $2.8-7 \times 10^{44}\ \rm erg\ s^{-1}$ range (see Table~\ref{tab:phys} for the individual values).

\begin{figure}
\begin{center}
\includegraphics[width=0.52\textwidth]{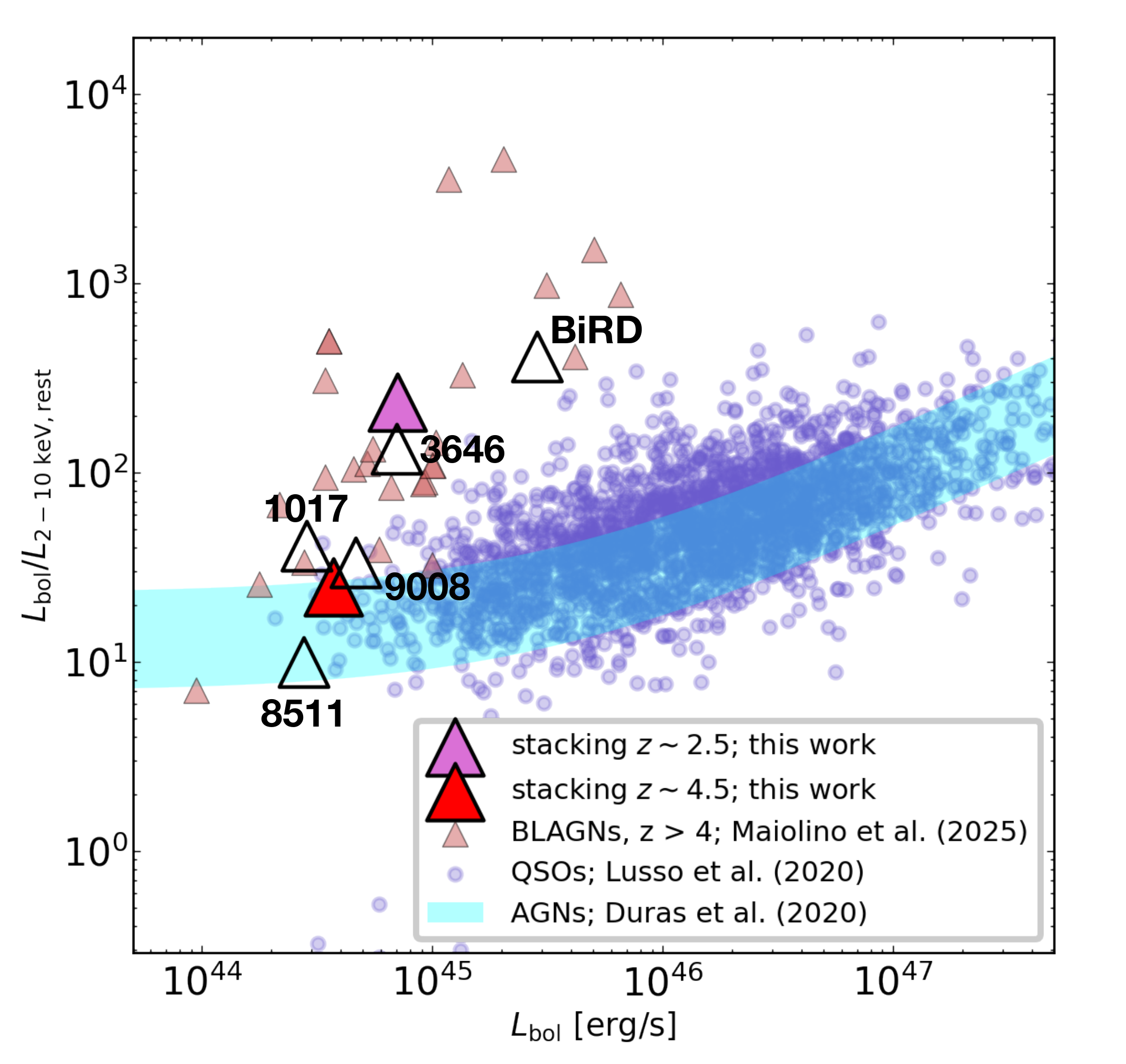}\\
\end{center}
\caption{X-ray bolometric correction $L_{\rm bol}/L_{\rm 2-10\ keV,rest}$ vs bolometric luminosity $L_{\rm bol}$. Lower limits to $L_{\rm bol}/L_{\rm 2-10\ keV,rest}$ for the BLAGNs in our sample are shown with empty triangles. We also show the lower limits corresponding to the stacked samples at $z\sim 2.4$ and $z\sim 4.5$.
The lower limits for the X-ray silent, BLAGNs at $z>4$ of \citet{maiolino24} are shown with indian red triangles.
For comparison, we show broad line quasars at various redshifts (slate blue filled circles; \citealt{lusso20}) and the relation found by \citet{duras20} for “classical” BLAGNs with its dispersion (cyan area).
The BLAGNs in our sample are undetected in the X-ray, with a bolometric correction up to a factor of $\sim 8$ (for BiRD $>100$) times larger than the expected value for standard AGNs.}
\label{fig:xbol}
\end{figure}

%X-RAY LUMINOSITY (INDIVIDUAL VALUES AND STACKING)
By selection, none of the sources in our sample of BLAGNs is detected in the X-ray $500\ \rm ks$ Chandra image of the \textit{J1030 field} \citep{nanni20, marchesi21}. Therefore, we derive $3\sigma$ upper limits to the X-ray luminosity adopting the same method used by \citet{loiacono25} (see their Sect. 5.1). For each source, we extracted background counts from annuli with inner radii 1.5\arcsec, 2.0\arcsec\ and 2.5\arcsec, for the 0.5-7 keV (full band, FB), 0.5-2 keV (soft, SB), 2-7 keV (hard, HB) band, respectively, and outer radii of 10\arcsec\ for all X-ray bands, and then computed the binomial no-source probability \citep{weisskopf07,vito19,nanni20}.
We then converted the counts into limits to the source count rates after correcting for the assumed aperture and for the small loss of effective area at the source's position.
By assuming no intrinsic absorption and a standard power-law spectrum with a photon index of $\Gamma=1.9$ (e.g., \citealt{peca21,signorini23}), we obtained for each source $3\sigma$ upper limits to the flux in the FB, SB, and HB, respectively. The soft band flux returns the most stringent constraint to the 2-10 keV rest-frame luminosity $L_{\rm 2-10\ keV, rest}$. We report the values of $L_{\rm 2-10\ keV,rest}$ in Table~\ref{tab:phys}. The X-ray bolometric correction $L_{\rm bol}/L_{\rm 2-10\ keV,rest}$ is in the range $\sim 10-130$ (while for BiRD it is $> 400$). These values are up to a factor of $\sim 4$ ($\sim 200$ for BiRD
) larger than the average expected value for standard AGNs (see Fig.\ref{fig:xbol}). This can be due either to strong obscuration or an intrinsically weak X-ray emission \citep{maiolino25, comastri26}. 

We also stacked the X-ray images at the source's location separately for the $z\sim 2.4$ and $z\sim 4.5$ samples. In this way we obtain more stringent upper limits to their median X-ray luminosity, which is $L_{\rm 2-10\ keV,rest} < 3.0 \times 10^{42}\ \rm erg\ s^{-1}$ at $z\sim 2.4$ and $L_{\rm 2-10\ keV,rest} < 1.5 \times 10^{43}\ \rm erg\ s^{-1}$ at $z\sim 4.5$. Assuming the median bolometric luminosity of the sources at $z\sim 2.5$ and $z\sim 4.5$, we find lower limits to the X-ray bolometric correction of $L_{\rm bol}/L_{\rm 2-10\ keV,rest} > 234$ and $L_{\rm bol}/L_{\rm 2-10\ keV,rest} > 25$, respectively. While the limit for the $z=4.5$ stack does not provide stringent information about the obscuration level, if the three objects at $z=2.4$ were standard AGN, it would take a column density of $N_{\rm H} \approx 10^{24}\ \rm cm^{-2}$, i.e. Compton thick or close to it, to bring the bolometric correction of their stacked emission at least a factor of 10 above the average value expected by \citet{duras20} for normal quasars.

%RADIO LUMINOSITY
None of the BLAGNs is detected in the radio image of the \textit{J1030 field} \citep{damato22}, which reaches an \textit{rms} of $2.5\rm\ \mu Jy$ at $1.4 \rm \ GHz$. Such non-detections imply upper limits in radio luminosity ($1.4\ \rm GHz$ rest frame) of $\sim 3.4 \times 10^{39}\ \rm \ergs$ at $z = 2.4$  and $\sim 1.3 \times 10^{40}\ \ergs$ at $z = 4.5$. Based on the typical emission level of radio-quiet AGN, these limits are not as stringent as the X-ray ones. Deeper radio observations are thus needed to verify if these sources are in fact radio weak or even silent, as shown by \citet{mazzolari24} for JWST-discovered BLAGNs.

% SUGLI UPPER LIMITS PER LA LBOL
%Si può anche tagliare dato che non lo mostriamo nel plot
%We also derived an upper limit for the bolometric luminosity based on the maximum permitted flux of the broad \Pag\ line. This value was computed by enhancing the flux of the broad \Pag\ component until the corresponding model had a $BIC$ higher than 10 with respect to the best model.

\subsection{Black hole masses}
\label{sub:bh}

We estimate the black hole mass using the relation of \citet{reines&volonteri15}, which is based on the \Ha\ luminosity and $FWHM$

{\small
\begin{equation}
    \log \left( \frac{M_{\rm BH}}{\rm M_{\odot}} \right) = 6.60 + 0.47 \log \left( \frac{L(\rm H \alpha)}{10^{42} \rm erg\ s^{-1}} \right) + 2.06 \log \left( \frac{FWHM(\rm H \alpha)}{1000\ \rm km\ s^{-1}} \right)
\label{eq:rv15}
\end{equation}
}

This relation has been used to estimate the black hole mass in several BLAGNs discovered with JWST (e.g., \citealt{matthee24, lin24, maiolino24}) and has a $0.5\ \rm dex$ scatter. Adopting a different calibration, i.e., \citet{greene&ho05}, lowers the black hole masses by a factor of $\sim 2$. These virial relations have been found to be adequate for LRDs, based on the direct measurement in \citet{juodzbalis26}.

As done in \citet{loiacono25}, for ID 1017 and ID 3646, we inferred the $FWHM(\rm H \alpha)$ from $FWHM(\rm Pa \gamma)$, assuming a ratio $\sim 1.57$ between the line widths. We estimated this ratio from the Rosetta Stone \citep{juodzbalis24}. We note that \citet{wang25} found a lower ratio ($FWHM(\rm H \alpha)/FWHM(\rm Pa \gamma) \sim 1.43$) in another well-studied LRD at cosmic noon (RUBIES-BLAGN-1). We account for this difference when calculating the errors on the black hole mass. 
Using Eq.~\ref{eq:rv15} we estimate a black hole mass $M_{\rm BH} = 1.7_{-1.1}^{+0.8} \times 10^7 \rm M_{\odot}$ and $M_{\rm BH} = 3.5^{+0.8}_{-1.7} \times 10^7 \rm M_{\odot}$ for ID 1017 and ID 3646, respectively. On the other hand, if we derive the black hole mass using the relation of \citet{landt13} (which has, however, one dex scatter), which is based on the broad \Pag\ emission and underlying continuum luminosity, we find $M_{\rm BH} \sim 1.2 \times 10^7 \rm M_{\odot}$ for ID 1017 and $M_{\rm BH} \sim 1.2 \times 10^7 \rm M_{\odot}$ for ID 3646. The black hole masses are well consistent within $1\sigma$ or $2\sigma$ for ID 1017 and ID 3646, respectively.
%when considering the $1\ \rm dex$ scatter of the relation of \citet{landt13}. 
In the rest of the work, we consider the \Ha -based black hole mass as the fiducial one to easily compare with other works that adopt the \citet{reines&volonteri15} calibration.

Finally, we directly estimate the black hole mass of ID 8511 and ID 9008 from their \Ha\ broad emission as $M_{\rm BH} = 1.6^{+0.8}_{-0.5} \times 10^7 \rm M_{\odot}$ and $M_{\rm BH} = 8.3^{+1.1}_{-0.9} \times 10^6 \rm M_{\odot}$ (see Table~\ref{tab:phys}).

\begin{table*}
\caption{Some physical and spectral quantities of the BLAGNs at $z\sim 2.4$ (ID 1017, ID 3646, and BiRD) and $z\sim 4.5$ (ID 8511 and ID 9008). We note that for the $z\sim 2.4$ sources, the \Ha\ luminosity was inferred from the \Pag\ one (see Sect.~\ref{sub:xrp}). We also show the BiRD values estimated by \citet{loiacono25}.}  
\label{tab:phys}      
\centering                               
\scalebox{0.9}{
\begin{tabular}{c c c c c c c}  
\hline
Source & $L$(\Ha) & $L_{\rm bol}$ &$L_X$ & $M_{\rm BH}$ & $\beta_{\rm UV}$ & $\beta_{\rm opt}$ \\[1mm] 
     & ($10^{42}\ \rm erg\ s^{-1}$) & ($10^{44}\ \rm erg\ s^{-1}$) & ($10^{42}\ \rm erg\ s^{-1}$) & ($10^7\ \rm M_{\odot}$) & &\\
\hline  
ID 1017 & $2^{+3}_{-1.7}$ & $3^{+4}_{-2}$ & $<7.0$ & $1.7_{-1.1}^{+0.8}$&$-1.01\pm0.22$&$0.00\pm0.17$ \\[1mm] 
ID 3646 & $5^{+3}_{-3}$ & $7^{+4}_{-4}$ & $< 5.3$ & $3.5^{+0.8}_{-1.7}$&$-0.62\pm0.42$&$-1.79\pm0.20$ \\[1mm]
BiRD & $26^{+3}_{-4}$ & $33^{+5}_{-5}$ & $< 7.0^*$ & $24_{-13}^{+10}$&$-0.87\pm0.17$&$2.71\pm0.19$ \\[1mm]
ID 8511 & $2.1^{+0.3}_{-0.3}$ & $2.8^{+0.4}_{-0.4}$ & $< 28 $ & $1.6_{-0.5}^{+0.8}$&$0.00\pm0.38$&$2.46\pm0.13$ \\[1mm] 
ID 9008 & $3.6^{+0.2}_{-0.2}$ & $4.7^{+0.2}_{-0.2}$ & $< 14 $ & $0.83_{-0.09}^{+0.10}$&$-1.22\pm0.10$&$1.88\pm0.08$ \\[1mm]  
\hline 
\end{tabular}}
\tablefoot{The errors are computed from the 16th and 84th percentiles of the fitted parameters in the MCMC fit. $^*$This value corrects an inaccuracy in the estimate reported by \cite{loiacono25} and is a factor of $\sim 2$ higher than the value given there.}
\end{table*}

\subsection{Bolometric luminosity function of LRDs at $z = 2.4$ and $z = 4.5$}
\label{sub:spaceden}

In Sect.~\ref{sub:blagn} we mention that our sources consist of three LRDs, one LBD, and one intermediate object (see also Sect.~\ref{sub:redvsblue}). However, we do not split our sample into sub-populations, and we use all the X-ray silent, red, and compact BLAGNs to derive the bolometric luminosity function (LF) of LRDs. We adopt this choice to compare with the results of other studies, which also derived the LF of LRDs without dividing their objects into sub-samples of LRDs and LBDs (e.g., \citealt{matthee24, taylor24, ma25, rinaldi26, kokorev24}).
%We do not split our sample into sub-populations (i.e., LRDs and LBDs; see Sect.~\ref{sub:redvsblue}), and we use the X-ray silent, red, and compact BLAGNs to derive the bolometric luminosity function (LF) of LRDs. A similar approach was also adopted by other studies, which derived the LF of LRDs at several wavelengths without dividing their objects into sub-samples (e.g., \citealt{matthee24, taylor24, ma25, rinaldi26, kokorev24}).
We compute the space density in two redshift bins ($z=2.4$ and $z=4.5$) using the $1/V_{\rm max}$ method \citep{schmidt68}. The number density $\Phi (L)$ of objects with $L$ luminosity is  

\begin{equation}
\label{eq:phi}
\Phi(L) = \frac{1}{\Delta \log L} \sum_{\rm i} \frac{w_{\rm i}}{V_{\rm max,i}},
\end{equation}

where $\Delta \log L$ is the width of the luminosity bin, $w_{\rm i} (\geq 1)$ are weights that account for possible incompleteness, and $V_{\rm max,i}$ is the maximum volume in which the $i$-th object would have been detected, i.e.,

\begin{equation}
\label{eq:vmax}
V_{\rm max,i} = f_\Omega \int_{z_{\rm min,i}}^{z_{\rm max,i}} \frac{dV_{\rm c}}{dz} dz ,
\end{equation}

where $dV_{\rm c}/dz$ is the comoving volume element and $f_\Omega$ is the fraction of the solid angle of the sky covered by the observations.

We computed the $1\sigma$-errors on $\Phi(L)$ following the approximations of \citet{gehrels86}. 
We use $z_{min} = 1.7$ and $z_{max} = 2.7$ for the $z\sim 2.4$ LRDs, while we adopt $z_{min} = 3.7$ and $z_{max} = 5.1$ for the $z\sim 4.5$ objects. At redshift lower (higher) than $z_{min}$ ($z_{max}$), the \Hei\ and \Pag\ (\Ha , for the $z\sim 4.5$ sample) lines fall outside the spectral range covered by the $F356W$ and grism $R$. 

Although some incompleteness may arise from the difficulty of detecting broad-line components (see, e.g., Sect.~\ref{sub:fit_birds}), we conservatively assume $w_i = 1$ for all of our targets.

The solid angle covered by the EIGER data is $\Omega_{\rm J1030} = 25\ \rm arcminutes^2$, which corresponds to $f_{\Omega} = 1.8 \times 10^{-7}$.
We use two 0.5 dex bins, centered at $\log(L_{\rm bol}/\rm erg\ s^{-1}) = (44.75, 45.25)$, for the $z\sim 2.4$ sample, populated by 2 and 1 sources, respectively, while we adopt one single 0.5 dex bin centered at $\log(L_{\rm bol}/\rm erg\ s^{-1}) = 44.5$ for the $z\sim 4.5$ sources.

The $\Phi(L_{\rm bol})$ values in the two redshift bins are reported in Table~\ref{tab:spacedens}.

We note that the space density of LRDs at $z\sim 4.5$ could be biased to higher values because of a possible clustering of the $z\sim 4.5$ BLAGNs in our sample, as their redshift separation is $\Delta z = 0.0015$ only (see Sect.~\ref{sub:blagn}). Nevertheless, our estimate is well consistent with the findings of other works at similar redshift and luminosity (see Sect.~\ref{sub:spacedendisc}).

\begin{table}
\caption{Bolometric luminosity function of LRDs at $z\sim 2.4$ and $z\sim 4.5$. We also report the number of sources per luminosity bin.}  
\label{tab:spacedens}      
\centering                               
\scalebox{0.8}{
\begin{tabular}{c c c}  
\hline
\multicolumn{3}{|c|}{$z\sim 2.4$} \\
\hline
$\log (L_{\rm bol}/\rm erg\ s^{-1})$ & $\Phi(L_{\rm bol})\ (10^{-5}\rm dex^{-1}\ Mpc^{-3} )$ & $N_{\rm sources}$\\[1mm] 
\hline  
44.75 & $4.5_{-2.9}^{+5.9}$ & 2\\[1mm] 
45.25 & $2.2_{-1.9}^{+5.2}$ & 1 \\[1mm] 
\hline
\multicolumn{3}{|c|}{$z\sim 4.5$} \\
\hline
$\log (L_{\rm bol}/\rm erg\ s^{-1})$ & $\Phi(L_{\rm bol})\ (10^{-5}\rm dex^{-1}\ Mpc^{-3} )$ & $N_{\rm sources}$\\[1mm] 
\hline
44.5 & $3.7_{-2.4}^{+4.9}$ & 2 \\[1mm] 
\hline 
\end{tabular}}
\tablefoot{The errors correspond to the $1\sigma$ uncertainties of \citet{gehrels86}.}
\end{table}

\begin{table}
\caption{Black hole mass function of LRDs at $z\sim 2.4$ and $z\sim 4.5$. We also report the number of sources per mass bin.}  
\label{tab:bhmf}      
\centering                               
\scalebox{0.8}{
\begin{tabular}{c c c}  
\hline
\multicolumn{3}{|c|}{$z\sim 2.4$} \\
\hline
$\log (M_{\rm BH}/\rm M_{\odot})$ & $\Phi(M_{\rm BH})\ (10^{-5}\rm dex^{-1}\ Mpc^{-3} )$ & $N_{\rm sources}$\\[1mm] 
\hline  
7.5 & $2.3_{-1.4}^{+3.0}$ & 2\\[1mm] 
8.5 & $1.1_{-0.9}^{+2.6}$ & 1 \\[1mm] 
\hline
\multicolumn{3}{|c|}{$z\sim 4.5$} \\
\hline
$\log (M_{\rm BH}/\rm M_{\odot})$ & $\Phi(M_{\rm BH})\ (10^{-5}\rm dex^{-1}\ Mpc^{-3} )$ & $N_{\rm sources}$\\[1mm]
\hline
7 & $3.7_{-2.4}^{+4.9}$ & 2 \\[1mm] 
\hline 
\end{tabular}}
\tablefoot{The errors correspond to the $1\sigma$ uncertainties of \citet{gehrels86}.}
\end{table}

\section{Discussion}
\label{sec:disc}

\subsection{Little red dots and little blue dots}
\label{sub:redvsblue}
\begin{figure}
\begin{center}
\includegraphics[width=0.48\textwidth]{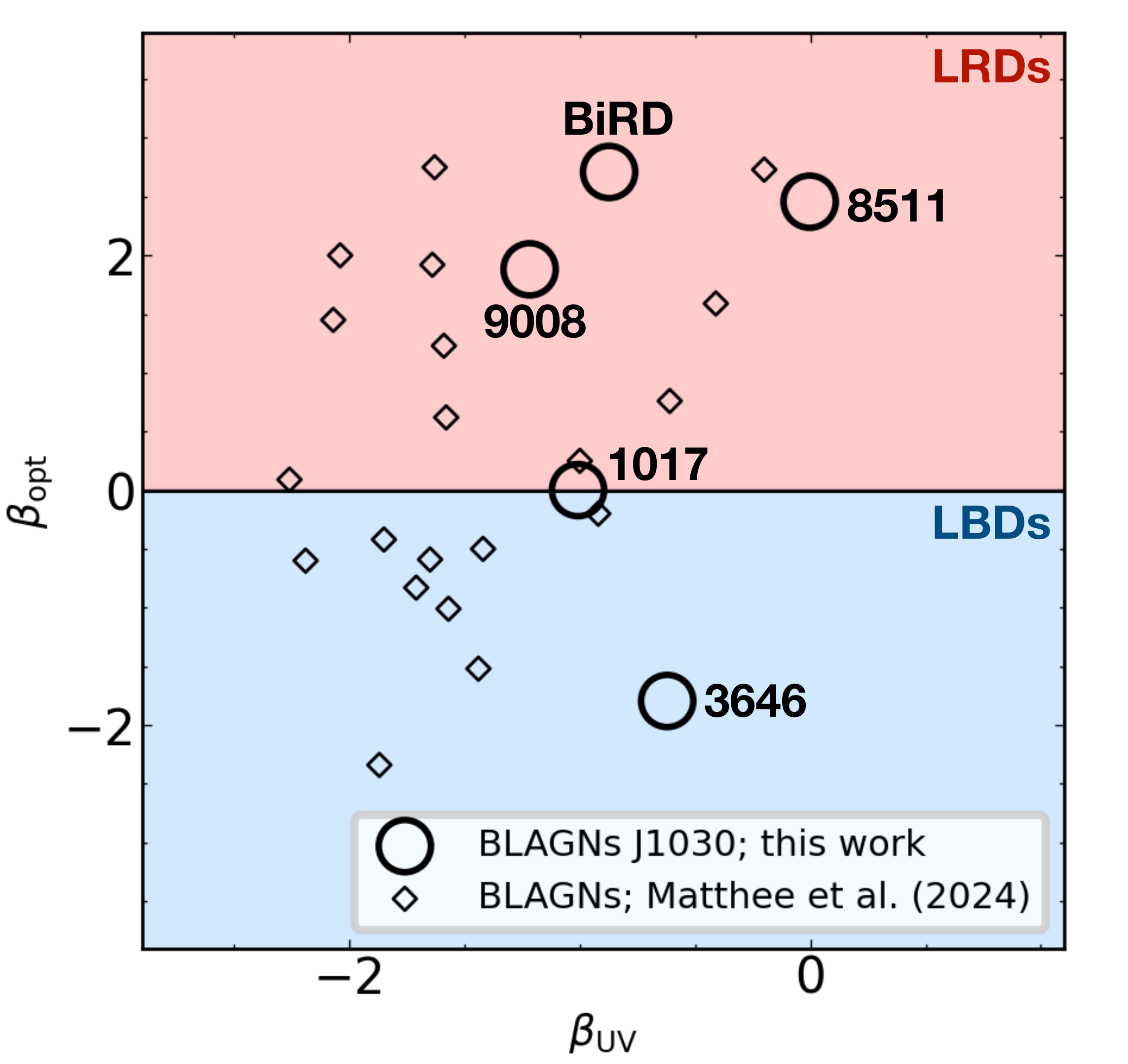}\\
\end{center}
\caption{UV slope $\beta_{\rm UV}$ vs optical slope $\beta_{\rm opt}$ for the X-ray silent BLAGNs in the J1030 field (circles). The red half-plane with $\beta_{\rm opt} > 0$ encloses the little red dots (LRDs), while little blue dots (LBDs) have $\beta_{\rm opt} < 0$ (blue half-plane). BiRD, ID 8511, and ID 9008 are clearly LRDs. On the other hand, ID 3646 has slopes consistent with LBDs. ID 1017 shows intermediate features between the two populations. We also show the sample of JWST-discovered BLAGNs of \citet{matthee24}, who used selection criteria very similar to ours.}
\label{fig:uvoptslopes}
\end{figure}

As shown in Fig.\ref{fig:uvoptslopes}, our selection criteria - i) point-like source; ii) presence of broad emission lines; iii) absence of X-ray detection (see Sect.~\ref{sub:sel}) - have returned three LRDs, one LBD, and one object (ID 1017) at the border between the two classes. We note that ID 8511 has $\beta_{\rm UV} > -0.37$ (i.e., the upper boundary of \citet{kocevski25} to define LRDs). This slope cut minimizes the contamination from dusty systems \citep{brazzini26, geris26}. As our sample is not affected by these contaminants, we relaxed this criterion and included ID 8511 in our sample. 

It is now interesting to compare the prevalence of LRDs in our sample with what is found in other JWST BLAGN samples, as the fractions of LRDs reported in different works scatter significantly. For example, taking the reported numbers at face value, some works indicate that LRDs may constitute only a minority, $\sim 30\%$, of the whole new population of BLAGNs discovered by JWST \citep{taylor24, brazzini26}, whereas other works find they are instead the vast majority (up to $\sim60-80\%$ of the population; \citealt{hviding25, zhuang25}, and, albeit with large uncertainties, even this work, where they are $\sim60-80\%$). What are the reasons for the reported differences, and, in turn, what is the true incidence of LRDs among JWST BLAGNs? As already noted in different works \citep{hainline25, taylor24, zhuang25}, the fraction of LRDs increases towards larger \Ha\ luminosities. On average, JWST BLAGN samples selected through NIRSpec spectroscopy reach lower luminosities (e.g. $\sim4\times 10^{41}$ erg s$^{-1}$ at $z=4$; \citealt{taylor24}) than those selected through NIRCam/WFSS ($\sim2\times 10^{42}$ erg s$^{-1}$; \citealt{matthee24, zhuang25}, this work), and is therefore reasonable to expect lower fractions in NIRSpec-selected samples. 
%\textbf{Also, LRDs are easier to select via photometry, due to the v-shaped SED, and so can be followed up spectroscopically more easily. On the other hand, LBDs have colors similar to those of normal galaxies/AGNs, and thus they are more difficult to be selected for follow-ups.}
In addition, we verified that a stronger effect is related to the velocity threshold adopted to define a broad line. For example, when increasing the line $FWHM$ threshold from $700$ to $1000$ and $1500$ \kms , the LRD fraction in the sample of \citet{taylor24} progressively increases from $34\pm6$\% to $39\pm7$\% and $50\pm8$\%. Also, it is worth noting that \citet{taylor24} only apply a compactness criterion to define LRDs, but not LBDs. As shown by \citet{hviding25}, whereas the vast majority ($\sim90$\%) of LRDs are compact, only $\sim60$\% of LBDs are compact. When summing compact and non-compact LBDs to compact LRDs, the fraction of LRDs with $FWHM > 1500$ \kms\ in \citet{hviding25} turns out to be $59\pm6$\%, very similar to the $50\pm8$\% in the \citet{taylor24} sample. In the WFSS sample by \citet{matthee24}, the LRD fraction goes from $55\pm11$\% to $59\pm12$\% when cutting at $FWHM>1000$ and $1500$ \kms , respectively, which is consistent with the results above.

To summarize, the LRD fraction increases towards higher line luminosities and width. The dependence on line width may be related to geometric effects consistent with recent modeling \citep{madau25}, suggesting that JWST BLAGNs are powered by Super Eddington flows and feature a highly asymmetric geometry. In these models, a geometrically thick accretion flow producing strongly anisotropic radiation is surrounded by a coplanar, equatorially concentrated BLR and, on larger scales, by a flattened dusty cloud distribution, both sharing a similar, modest ($\sim 10-20$\%) covering factor. In this scenario, LRDs would be the edge-on analogs of face-on LBDs, and hence would be preferentially selected at large line widths. Instead, it would be difficult to reconcile the prevalence of LRDs at large \Ha\ luminosities if the covering factor is low. In fact, in (edge-on) LRDs the \Ha\ luminosity would be more suppressed by dust than in (face-on) LBDs, and only a large dust covering factor would be able to reproduce the observed LRDs dominance, e.g. $\sim$70\% for $ L(\rm H\alpha)>10^{43}\ \ergs$ in the samples of \citet{taylor24} and \citet{matthee24}. This argument could suggest that LRDs are intrinsically more abundant than LBDs at a given bolometric (line) luminosity, and that the smaller fraction of LRDs observed at low luminosities is primarily the result of selection effects, as extinction in edge-on systems can suppress faint emission lines below the detection threshold.

\subsection{Bolometric luminosity function of LRDs and other AGN populations}
\label{sub:spacedendisc}
\begin{figure*}
\begin{center}
\includegraphics[width=0.98\textwidth]{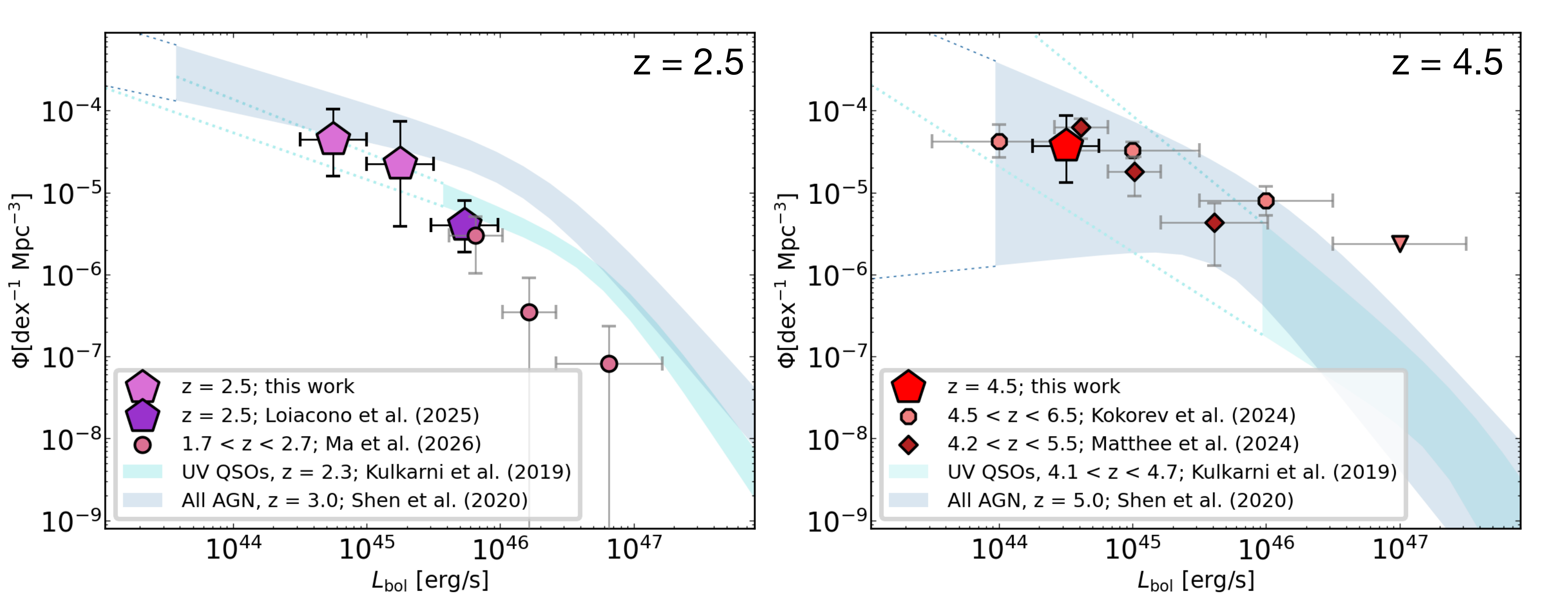}\\
\end{center}
\caption{Bolometric LF of LRDs at $z\sim 2.4$ and $z\sim 4.5$. \textit{Left panel}: we show with orchid pentagons the abundance of LRDs estimated from the \textit{J1030 field}. The estimate from \citet{loiacono25} is shown with a purple pentagon. The LF from \citet{ma25} is shown with filled circles and includes their complete bins after converting the de-extinguished $M_{5500}$ magnitude to bolometric luminosity (see the text for the details).
We also show the bolometric LF derived at similar redshifts for the entire AGN population (steel blue; \citealt{shen20}) and for UV-selected, unobscured AGNs (pale turquoise; \citealt{kulkarni19}). The shaded areas correspond to the 16th and 84th percentiles of the bolometric LFs. The dotted lines enclose the bolometric luminosity range for which the abundance of AGNs was extrapolated from data at larger $L_{\rm bol}$. \textit{Right panel:} the red pentagon shows the space density of LRDs based on the \textit{J1030 field}. We also show the estimate of \citet{kokorev24} and \citet{matthee24} derived at similar redshifts (the upper limits from \citealt{kokorev24} are shown with inverted triangles). The LFs of quasars and UV-bright AGNs are also shown.}
\label{fig:lf}
\end{figure*}

We show the bolometric LF of LRDs at $z\sim 2.5$ and $z\sim 4.5$ (see Sect.~\ref{sub:spaceden}) in Fig.~\ref{fig:lf}.

% COMPARISON WITH QUASARS
First, we compare the abundance of LRDs with that of UV-selected quasars (\citealt{kulkarni19}; see also \citealt{niida20}) and the whole AGN population based on pre-JWST studies \citep{shen20}. The luminosity function of \citet{kulkarni19} was computed at rest-frame 1450 \AA\ absolute magnitude and considers unobscured AGNs up to $z\sim 7.5$. We convert their UV to bolometric luminosity adopting the bolometric correction of $L_{\rm bol}/L_{1450} = 4.0$ from \citet{richards06}, consistent with \citet{runnoe12}. 
\citet{shen20} considered AGN samples selected from the IR to the X-rays and, based on SED fitting, derived bolometric and extinction corrections for all AGNs in their samples. Because of the adopted selection, their bolometric LFs likely include both unobscured and obscured AGNs (i.e., what was considered as the "whole" AGN population before JWST) up to $z\sim 7$. 
In Fig.~\ref{fig:lf} we show for both works the $L_{\rm bol}$ range for which the LF was directly measured based on the observed samples (shaded area) or only extrapolated based on higher luminosity data (dotted lines).
The shaded areas were conservatively computed by considering all the possible combinations of the $1\sigma$ errors of the best-fit parameters of \citet{shen20} and \citet{kulkarni19}.

At $z\sim 2.4$, we see that the abundance of LRDs is only a factor of $\sim 2-2.4$ lower than that of all AGNs with comparable bolometric luminosity (Fig.~\ref{fig:lf}, left panel). Their number density is fully consistent with the inferred abundance of UV-selected quasars with $L_{\rm bol} \sim 10^{45}\ \rm erg\ s^{-1}$.
This means that the abundance of LRDs is still significant at cosmic noon, being comparable with that of quasars with similar bolometric luminosity.

At $z\sim 4.5$, the number density of LRDs is consistent within the errors with that of all pre-JWST AGNs of \citet{shen20} (Fig.~\ref{fig:lf}, right panel).

% COMPARISON WITH OTHER LRDs
We then compare the number density of LRDs with that estimated for the same population from other studies at similar redshifts, namely \citet{loiacono25} and \citet{ma25} (at $z\sim 2.4$),
\citet{kokorev24} and \citet{matthee24} (at $z\sim 4.5$). We note that \citet{matthee24} compute the number density of LRDs at $4.2 < z < 5.5$ as a function of the \Ha\ luminosity. We converted their \Ha\ luminosities to the bolometric ones adopting Eq.~\ref{eq:stern}, as done for the objects in this work (see Sect.~\ref{sub:xrp}). On the other hand, \citet{ma25} derived the abundance of LRDs at $1.7 < z < 2.7$ as a function of the absolute magnitude at $5500$ \AA\ ($M_{5500}$) using ground-based observations.
We caution the reader that the absolute magnitude of \citet{ma25} is not corrected for dust attenuation. Therefore, to compare with our work, we de-estinguished the $5500$ \AA\ magnitude by assuming an attenuation $A_{\rm V} = 1.6$. This is the average value of the sample of \citet{kokorev24} estimated with SED fitting and assuming an AGN template. We then converted the $5500$ \AA\ de-extincted luminosities to bolometric ones using a bolometric correction of $\sim 10$ from \citet{richards06}. We note that at $1.7< z <2.7$ the luminosity function of \citet{ma25} is complete up to $M_{5500} = -21$ and thus we exclude their first three bins from the comparison (see their Table~1). The resulting bolometric luminosities are $\log (L_{\rm bol}/\rm erg\ s^{-1}) \gtrsim 45.75$ (Fig.~\ref{fig:lf}, left).

Compared to \citet{loiacono25}, we are sampling LRDs at $z \sim 2.4$ with lower bolometric luminosity, finding a decline of the space density between $\log (L_{\rm bol}/\rm erg\ s^{-1}) = 45.25$ and $\log (L_{\rm bol}/\rm erg\ s^{-1}) = 45.75$.
%We possibly find a decline of the space density between $\log (L_{\rm bol}/\rm erg\ s^{-1}) = 45.25$ and $\log (L_{\rm bol}/\rm erg\ s^{-1}) = 45.75$ by a factor of $\sim 5.6$, when considering the central values. 
%We note, however, that BiRD is included in the estimate of \citet{loiacono25} too.
%If we recompute their space density after having excluded BiRD and recalculating the comoving volume, the decrease of the space density in the same luminosity bins becomes a factor of $\sim 7.6$.

%NOW COMPARE WITH MA+
From Fig.~\ref{fig:lf} (left) it is clear that \citet{ma25} are sampling the bright-end of the LF. Interestingly, at $\log (L_{\rm bol}/\rm erg\ s^{-1}) = 45.75$, i.e., the luminosity bin for which both the JWST- and ground-based estimates are available, the space densities from \citet{loiacono25} and \citet{ma25} are perfectly in agreement. 

% COMPARISON AT HIGH REDSHIFT
At $z\sim 4.5$, our estimate is nicely in agreement with the points from \citet{matthee24} and \citet{kokorev24} at similar luminosity and redshift (Fig.~\ref{fig:lf}, right). 
\citet{matthee24} selected LRDs from EIGER (excluding the \textit{J1030 field}, which had not yet been observed at that time) and FRESCO \citep{oesch23} programs based on their broad line emission over an area of $\sim 230\ \rm arcmin^2$, i.e., a factor $\times \sim 8.5$  larger than our field. We thus expect to sample only the more abundant low-luminosity end of their LF because of the smaller area of our field.

%HERE discussion about the extinction?

%Compared to \citet{matthee24}, who selected LRDs based on their broad line emission over an area a factor $\times \sim 4$ CHECK larger than our field,  

%DISCUSS WHY IT IS DIFFICULT TO COMPARE WITH WORKS THAT USE THE UV LUMINOSITY HERE OR IN THE SECTION ON THE EVOLUTION

\subsection{Active black hole mass function}
\label{sub:bhmassfunc}
%BHMF
\begin{figure*}
\begin{center}
\includegraphics[width=0.98\textwidth]{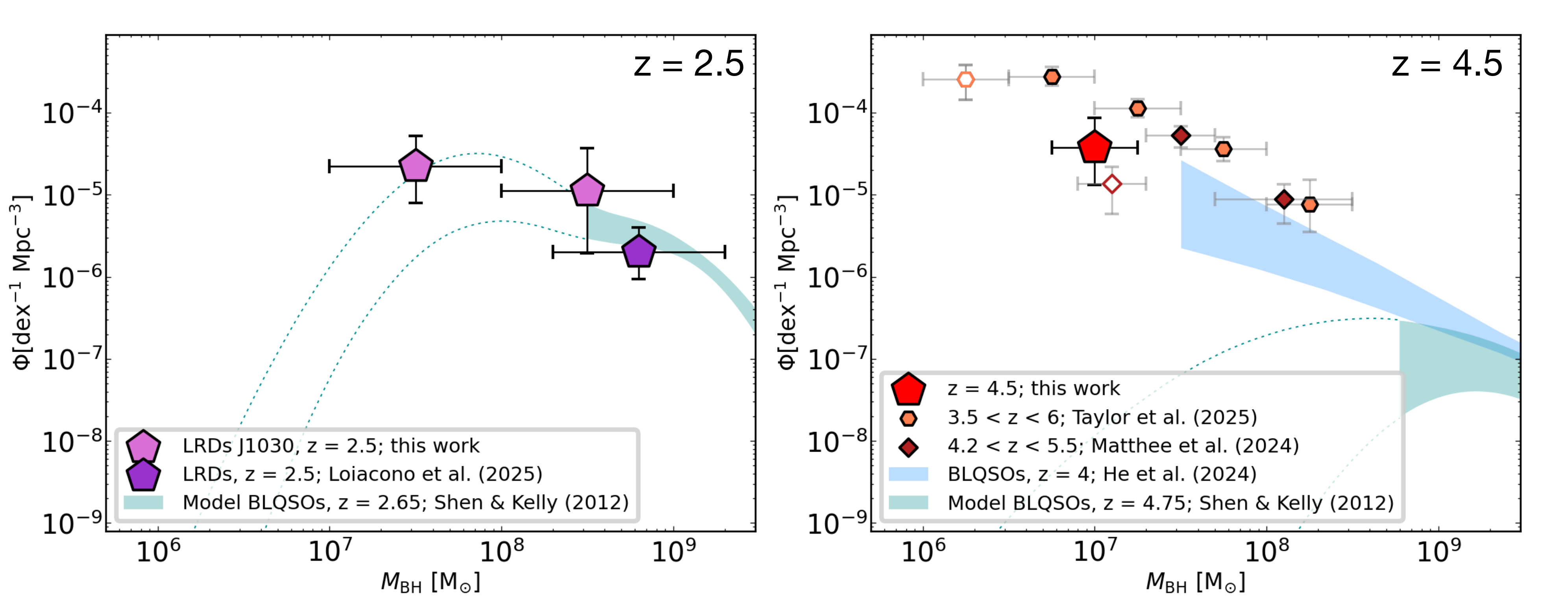}\\
\end{center}
\caption{Active black hole mass function (BHMF) of LRDs at $z\sim 2.4$ and $z\sim 4.5$. \textit{Left panel}: we show with orchid pentagons the abundance of LRDs estimated from the \textit{J1030 field}. The estimate from \citet{loiacono25} is shown with a purple pentagon. We also show the predicted active BHMF from broad line quasars at similar redshift (dark cyan; \citealt{shenkelly12})
The shaded areas correspond to the 16th and 84th percentiles. The dotted lines enclose the black hole mass range for which the abundance of broad line quasars was inferred from data with larger $M_{\rm BH}$. \textit{Right panel:} the red pentagon shows the space density of LRDs at $M_{\rm BH} = 10^7\ \rm M_{\odot}$ based on the \textit{J1030 field}. We also show the predicted active BHMF from broad line quasars at a similar redshift (dark cyan; \citealt{shenkelly12}). In addition, we show the best-fit from \citet{he24} to low-luminosity quasars, corrected for incompleteness. 
Finally, we show the active BHMF of LRDs at higher redshifts from \citet{matthee24} (dark red diamonds) and \citet{taylor24} (coral hexagons). We show with empty symbols the points from \citet{matthee24} and \citet{taylor24} affected by incompleteness.}
\label{fig:bhmf}
\end{figure*}
We study the space density of LRDs as a function of the black hole mass. We note that the active black hole mass functions (BHMFs) are, in general, difficult to derive due to selection biases \citep{matthee24, taylor24}. Therefore, we consider our measurement as tentative. 

At $z\sim 2.4$, the black hole masses of ID 1017, ID 3646, and BiRD span between $\sim 1.6\times 10^7 - 2.4 \times 10^8\ \rm M_{\odot}$ and thus we split the sources over two mass bins of 1 dex centered at $\log(M_{\rm BH}/\rm M_{\odot}) = (7.5, 8.5)$, populated with two and one sources, respectively. We divide the space density determined in Sect.~\ref{sub:spaceden} by a factor of 2 to take into account the 1 dex bin size. On the other hand, at $z\sim 4.5$, we use a 0.5 dex bin centered at $\log(M_{\rm BH}/\rm M_{\odot}) = 7$. The BHMFs are reported in Table~\ref{tab:bhmf} and shown in Fig.~\ref{fig:bhmf}.

%INCOMPLETENESS AT z = 4.5 (AND ALSO AT z = 2.5)
We note that the lowest mass bin at $z\sim 2.4$ and the mass bin at $z\sim 4.5$ possibly suffer from incompleteness because of the $FWHM$ threshold that we adopt to define a broad line emitter (see Sect.~\ref{sub:blagn}). Indeed, we are possibly missing sources with $M_{\rm BH} \sim 10^{7.5}$ \Msun\ and $M_{\rm BH} \sim 10^{7}$ \Msun , depending on the redshift bin, but with $FWHM < 1000$ \kms .   

%COMPARISON WITH QUASARS
Compared to quasars, at $z\sim 2.4$ the abundance of LRDs (plum pentagons in Fig.~\ref{fig:bhmf}, left) is consistent with the number density of BLQSOs inferred by \citet{shenkelly12}. 

At $z\sim 4.5$ LRDs with $M_{\rm BH} \sim 10^{7}$ \Msun\ (red pentagon, Fig.~\ref{fig:bhmf}, right) are significantly more abundant than quasars, according to the predicted values by \citet{shenkelly12}. We note however that \citet{shenkelly12} sample black holes with $M_{\rm BH} \gtrsim 6 \times 10^{8}$ \Msun , which makes the extrapolated abundance at $M_{\rm BH} \sim 10^{7}$ \Msun\ very uncertain. We also compare our estimate with the best-fit BHMF of \citet{he24} (corrected for incompleteness), which includes measurements from BLAGNs with black hole masses down to $M_{\rm BH} \sim 3 \times  10^{7}$ \Msun\ at slightly lower redshift ($3.50 < z < 4.25$), based on SDSS and the Subaru Hyper-Supreme Cam data.
Compared to \citet{he24}, our estimate of the BHMF samples lower black hole masses and appears a factor of $\sim 1.8\times$ larger than a possible extrapolation of their best-fit at lower masses (yet consistent within the $1\sigma$ uncertainties).  

%COMPARISON WITH OTHER LRDs STUDIES
We compare our estimates with the BHMFs of LRDs from other studies at $z\sim 2.5$ \citep{loiacono25} and $z\sim 4.5$ \citep{matthee24, taylor24}. We do not apply any scaling factor to their mass bins, as these works use the calibration of \citet{reines&volonteri15} to derive black hole masses. We note that, as for our measurement at $z\sim 4.5$, neither \citet{matthee24} nor \citet{taylor24} apply the extinction correction to the \Ha\ luminosity.

At $z\sim 2.5$, our BHMF in the largest mass bin is fully consistent with the value of \citet{loiacono25}, which samples black hole masses that are larger on average. However, we note that the two estimates are not independent as BiRD is included in the BHMF of \citet{loiacono25} as well.
%We compare our estimate at $z\sim 4.5$ with the BHMF of LRDs from other studies at similar redshift \citep{matthee24, taylor24}. 
%We do not apply any scaling factor to their mass bins, as both studies used the calibration of \citet{reines&volonteri15} to derive black hole masses. We note that, as in our case, none of these works apply extinction correction to the \Ha\ luminosity.

At $z\sim 4.5$, our estimate is nicely consistent with the lowest mass bin of \citet{matthee24}, which also suffers from incompleteness (empty diamond in Fig.~\ref{fig:bhmf}, right). On the other hand, the values of \citet{taylor24} at $6 \times 10^6 - 2 \times 10^7\ \rm M_{\odot}$ are a factor of $\sim  3-7$ larger than our point. This is likely because \citet{taylor24} do not apply any $FWHM$ cut to the \Ha\ emitters in their sample, which mitigates incompleteness effects. 
Besides, as discussed in Sect.~\ref{sub:redvsblue}, at low $FWHM$, the selection of \citet{taylor24} is more sensitive to LBDs than LRDs, compared to our sample. When considering LRDs only, as for our $z\sim 4.5$ bin, their points are only a factor of $\sim 2$ larger than our estimate.
%This is likely due to two effects. On one hand, \citet{taylor24} does not apply any $FWHM$ cut to the \Ha\ emitters in their sample, which mitigates any incompleteness effect. Second, \citet{taylor24} does not apply any cut in the UV slopes of their targets, thus including objects that are not LRDs. Over their bins with $M_{\rm BH} \sim 6 \times 10^6 - 2 \times 10^7\ \rm M_{\odot}$, the LRDs fraction is $\sim 25 - 65\%$ of the total. 
%As discussed in Sect.~\ref{sub:redvsblue}, the selection of \citet{taylor24} is indeed more sensitive to LBDs than LRDs, compared to our sample. When considering LRDs only, their points are only a factor of $\sim 2$ larger than our estimate.
%COMMENTO D'INSIEME
%Overall, our BHMFs suggest that at $z\sim 2.4$ the abundance of LRDs is comparable with that of quasars with similar BH masses, while at $z\sim 4.5$ the comparison with other BLAGNs is tricky due to the mass limits of current surveys.  
\subsection{Evolution of the number density of LRDs down to cosmic noon}
\label{sub:evolution}

\begin{figure*}
\begin{center}
\includegraphics[width=0.88\textwidth]{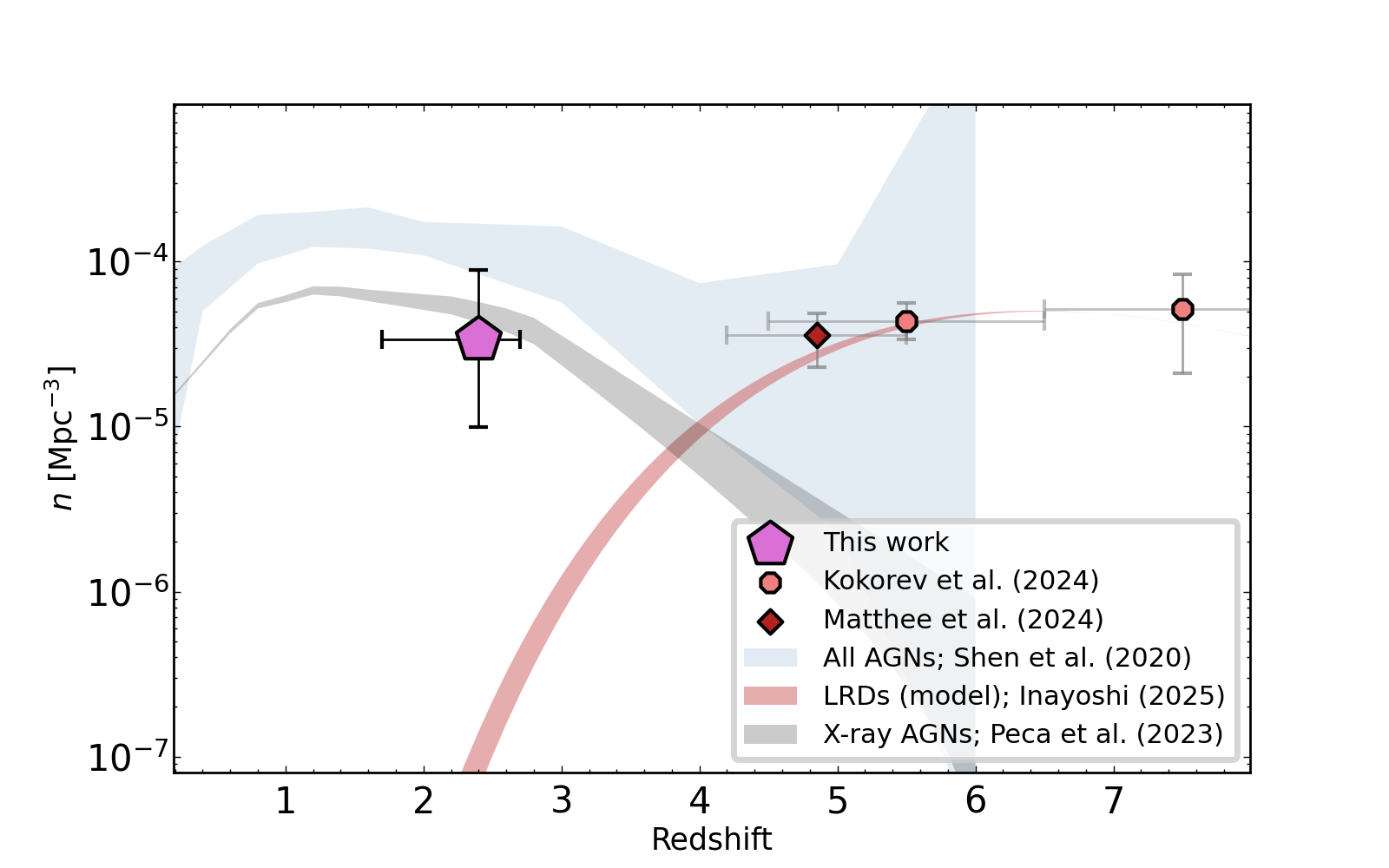}\\
\end{center}
\caption{Abundance of LRDs with $L_{\rm bol} \gtrsim 3 \times 10^{44}\ \rm erg\ s^{-1}$ as a function of redshift. We compare our estimate at $z \sim 2.4$ (orchid pentagon) with the number density of LRDs from \citet{matthee24} and \citet{kokorev24} over the same integration limits. We show the abundance of classical pre-JWST AGNs \citep{shen20} and X-ray selected AGNs \citep{peca23} integrated over the same bolometric luminosity range. At cosmic noon, the abundance of LRDs is a factor of $\sim 350$ higher than the predictions by \citet{inayoshi25}. The shaded areas correspond to the $1\sigma$ errors.}
\label{fig:evolut}
\end{figure*}
%Integration
We integrate the LFs determined in Sect.~\ref{sub:spaceden} to derive the number density of LRDs with $L_{\rm bol} \gtrsim 3 \times 10^{44}\ \rm erg\ s^{-1}$ (i.e., the minimum luminosity covered by our bins). We consider the $z\sim 2.4$ points only, as at $z\sim 4.5$ we have only one luminosity bin. However, the latter is consistent with other estimates at similar redshifts \citep{matthee24}. When summing our measurements down to that luminosity limit, the number density of LRDs is $n = 3.4^{+5.6}_{-2.4} \times 10^{-5}\ \rm Mpc^{-3}$. 
%This value is consistent with what we would obtain by performing a Schechter fit to the LF at $z\sim 2.4$, after including a third luminosity bin at $\log (L/\rm erg\ s^{-1}) = 45.75$, which considers the more luminous LRDs at cosmic noon "Rosetta stone" \citep{juodzbalis24} and RUBIES-BLAGN-1 (\citealt{wang25}; see \citealt{loiacono25}).
A similar value is obtained when fitting a Schechter function to all LF data points at $z \sim 2.4$ shown in Fig.~\ref{fig:lf} (left), including those in the highest luminosity bins from \citet{ma25}, and then integrating it down to the same $L_{\rm bol}$ limit.

To compare our estimate with the abundance of LRDs at higher redshifts, we integrate the bolometric LFs from \citet{matthee24} and \citet{kokorev24} over the same integration limits (i.e., excluding the first bolometric luminosity bin of \citealt{kokorev24}). We note that \citet{matthee24} selected their sample based on the broad ($FWHM > 1000$ \kms) \Ha\ emission (see also Sect.~\ref{sub:spacedendisc}). The resulting sources have a compact morphology, show red $F200W-F356W$ colors, and are undetected in the X-ray. In addition, based on their UV and optical slopes, the sample of \citet{matthee24} includes both LRDs and LBDs (see Fig.~\ref{fig:uvoptslopes}). Therefore, their selection is consistent with ours.
\citet{kokorev24} instead selected their LRDs candidates photometrically, based on their red colors and compactness. Thus, the missing information about the presence of broad lines and X-ray emission makes the comparison between our and their estimate of the number density uncertain.  

In Fig.~\ref{fig:evolut} we show the abundance of LRDs with $L_{\rm bol} \gtrsim 3 \times 10^{44}\ \rm erg\ s^{-1}$ as a function of redshift. According to our estimate, the number density of LRDs does not significantly evolve from $z \sim 7$ down to $z \sim 2.4$.

Using photometrically selected LRDs, \citet{ma25} find at $1.7 < z < 2.7$ an abundance of LRDs as low as $n \sim (2.1 \pm 1.1) \times 10^{-6}\ \rm Mpc^{-3}$. However, as we discuss in Sect.~\ref{sub:spacedendisc}, \citet{ma25} are sampling the bright-end ($L_{\rm bol} \gtrsim 6.5 \times 10^{45}\ \ergs$) of the bolometric LF (Fig.~\ref{fig:lf}, left). This explains why their integrated LF gives a number density a factor of $\sim 10$ lower than our estimate. 

The number density we derive at $z\sim 2.4$ is consistent with the value found by \citet{rinaldi26} at similar redshifts using photometrically selected LRDs. However, \citet{rinaldi26} consider absolute UV luminosities in their integration (down to $M_{\rm UV} = -18.5$). We thus caution the reader that our comparison is only qualitative because of the uncertainty in the conversion of UV magnitudes into AGN bolometric luminosity in LRDs (this is why we refrain to show the \citealt{rinaldi26} data-points in Fig.~\ref{fig:evolut}), as the nature of the UV emission in these objects is still debated. For example, the detection of an extended UV component in a fraction of LRDs \citep{rinaldi25b} leaves room for a significant contribution from stellar or scattered nuclear emission, preventing in any case a careful assessment of the intrinsic AGN power. 
%Indeed, if we convert the $UV$ absolute magnitude $M_{\rm UV} = -18.5$ (i.e, the integration limit used by \citealt{rinaldi26}) to bolometric luminosity, assuming the same attenuation used for \citet{ma25} and a bolometric correction of $\sim 4$ at 1450 \AA\ \citep{richards06}, we find that \citet{rinaldi25} are integrating the LRDs number density for $L_{\rm bol} \gtrsim 4 \times 10^{44}\ \rm erg\ s^{-1}$, which is very close to our integration limit. However, we caution the reader that this is a qualitative estimate only of the bolometric luminosity, as the nature of the UV emission (i.e., stars or accretion disk) in LRDs is still poorly understood. 

%CAVEAT DELL'ESTINZIONE
A possible caveat to discuss about the LRD number density at $z \sim 5$ is that the bolometric luminosities for the sample of \citet{matthee24} ware not corrected for dust extinction (see also Sect.~\ref{sub:spaceden}). This means that the number density that we computed from \citet{matthee24}'s LF could refer to objects with higher intrinsic bolometric luminosity than the $z \sim 2.4$ ones.
Therefore, we attempt to estimate how applying an extinction correction would impact their estimate of the LRDs number density at $z \sim 5$. Assuming a Small Magellanic Cloud extinction law\footnote{We note that similar results are obtained using different extinction laws.} \citep{prevot84} and an attenuation of $A_{\rm V} = 1.6$ from \citet{kokorev24}, we estimate the observed bolometric luminosity that would correspond to an intrinsic $L_{\rm bol} \gtrsim 3 \times 10^{44}\ \rm erg\ s^{-1}$. We thus derive the corresponding values of the LF by fitting the \citet{matthee24} points with a power-law. The resulting number density is a factor of $\sim 4.6$ higher than the current estimate. This would imply a decline in the number density of LRDs between cosmic dawn and cosmic noon by a factor of $\sim 5$ at most, which would not affect our conclusions significantly.  
   
%COMPARISON WITH LRDs MODELS AND IMPLICATIONS
We also compare our estimate with the model by \citet{inayoshi25}, which was tailored to reproduce the redshift distribution of the \citet{kocevski25} LRD sample (i.e. 341 photometrically selected objects in the redshift range $z \sim 2-11$). This model describes well the abundance of LRDs at $z \sim 4-7$ and predicts that the number density of LRDs drops at $z < 4$ (red shaded area in Fig.~\ref{fig:evolut}). We note that the model considers LRDs with absolute UV magnitude $ -22 < M_{\rm UV} \ [\rm mag] < -16$ (but the trend also persists considering the objects with $M_{\rm UV} < -18\ \rm mag$ only) rather than bolometric luminosity. Thus, also in this case, we can only do a qualitative comparison. The abundance of LRDs we measure at $z = 2.4$ is a factor of $\sim 350$ larger than the model predictions.

%MAYBE AT THE END OF THE SECTION
According to the \citet{inayoshi25} model, LRDs may represent the early, rapid stages of growth of black hole (heavy) seeds with mass $M_{\rm BH} \sim 10^4-10^5$ \Msun . The formation of black hole seeds is expected to be suppressed at $z < 5$, as metals diffuse in the interstellar medium, producing fragmentation and cooling of clouds, which prevent them from directly collapsing into black holes \citep{bellovary11}. Our findings may suggest that the formation of black hole seeds can still be efficient at epochs as recent as cosmic noon. Alternatively, our result may simply hint that LRDs, rather than being the very first accretion episodes of growing seeds, are more likely a high-accretion phase in already mature black holes. Additional studies will be necessary to discriminate between the two scenarios.
%This result confirms our previous findings \citep{loiacono25}.

\subsection{LRDs and other AGN populations at cosmic noon}
\label{sub:lrdsvsagn}

%COMPARISON WITH QUASARS, OUR ESTIMATE MA ANCHE ANDAMENTO GENERALE
We compare the abundance of LRDs at cosmic noon with that of AGNs based on multi-wavelength (bolometric) selection (light blue area in Fig.~\ref{fig:evolut}; \citealt{shen20}) or X-ray selection (gray shaded area; \citealt{peca23}). We obtained the curves by integrating the AGN LFs at various redshifts for the same integration limit of LRDs ($L_{\rm bol} \gtrsim 3 \times 10^{44}\ \rm erg\ s^{-1}$) and considering the $1\sigma$ errors. For the X-ray LF by \citet{peca23}, we converted from $2-10\ \rm keV$ to bolometric luminosities using the relation of \citet{duras20}.
The difference between the AGN space density derived using the bolometric LF of \citet{shen20} and the X-ray LF of \citet{peca23} is about a
factor of $2-3$ going from $z = 0$ to $z = 4$. This is broadly consistent with the expectations of synthesis models of the cosmic X-ray background \citep{gilli07, ananna19} suggesting that the abundance of heavily obscured, Compton-thick AGNs ($N_{\rm H} > 10^{24}\ \rm cm^{-2}$) that are largely missed by X-ray surveys, is comparable to that of less obscured AGNs whose abundance is instead recorded in the X-ray luminosity functions.

At $z \sim 2.4$ the abundance of LRDs with $L_{\rm bol} \gtrsim 3 \times 10^{44}\ \rm erg\ s^{-1}$ is a factor $\sim 3.6$ lower than that of all AGNs. On the other hand, the number density of LRDs is within the errors of X-ray selected AGNs with comparable bolometric luminosity. 

%INTERPRETATION: COMPTON THICK ET CETERA
It is possible that LRDs are part of the population of Compton-thick AGNs, whose contribution to the total AGN census is still poorly understood. 
For example, if the large equivalent widths that are measured in the broad \Ha\ lines of LRDs are due to a large covering factor of the clouds in the BLR (significantly larger than in standard AGN), then the same clouds would be sufficiently dense to produce Compton-thick absorption \citep{maiolino25}. 

However, absorption by Compton-thick gas perhaps is not the only reason for the lack of X-ray emission in LRDs, which might be instead intrinsically X-ray weak \citep{tortosa26}. Several LRDs have line widths $< 2000$ \kms\ and can thus be classified as narrow-line Seyfert 1 galaxies, which are known to accrete at high rates and to have very steep X-ray spectra \citep{iwasawa24}. Such steep spectra, would indeed imply low rates of X-ray photons at rest-frame energies above a few keV, i.e. the ones sampled in high-z LRDs, explaining in turn why these objects may have escaped X-ray detection \citep{maiolino25}. 
%Accretion at high rates can further explain their non-detection, even if other mechanisms such as strong obscuration, a combination of intrinsic suppression and absorption of the corona, and errors in the black hole mass estimates cannot be excluded \citep{tortosa26}.

Recently, \citet{sneppen26} proposed a scenario in which both obscuration by thick gas ($N_{\rm H} \sim 10^{25}\ \rm cm^{-2}$) and high accretion rates are needed simultaneously to explain the observed X-ray weakness of LRDs.
From the picture portrayed above, it is thus difficult to quantify to what extent LRDs contribute to the population of Compton-thick AGNs, with the latter possibly having a substantial role in the accretion history of supermassive black holes \citep{peca23}. 

%FRASE CONCLUSIVA (PRENDERE EVENTUALMENTE IL PARAGRAFETTO SOPRA)
%Therefore, our study suggests that there is no evolution in the number density of LRDs at least up to cosmic noon and that this population at $z\sim 2.4$ can be as abundant as coeval X-ray selected AGNs with comparable bolometric luminosity.
\subsection{The fate of LRDs in the local universe}
\label{sub:lrdsvslocal}

% COMPARISON WITH LOCAL STUDIES AT FACE VALUES
According to some studies, the number density of LRDs in the local universe becomes as low as $n \sim (1.6 - 5) \times 10^{-9}\ \rm Mpc^{-3}$ \citep{lin26, park26}. This implies a drop by a factor of $\sim 10^4$ compared to our estimate at cosmic noon. 

% SAMPLE SELECTIONS
A fair comparison between our measurement and those in the local universe is not obvious due to the different sample selections. 
%\citet{lin26} selected their objects from the SDSS DR17 spectroscopic database based on their 'v-shape' (i.e., including LRDs only), removed sources with strong \Nii\ and \Mgii\ emission, as possible type-I AGN contaminants, and required an \Oiii\ equivalent width $> 10 \AA$ to reduce the contamination from galaxies. The objects are compact on a scale of $1.8$\arcsec (corresponding to $3.3$ kpc at $z\sim 0.1$), and their final sample includes three objects at $z\sim 1$ selected as the most representative LRDs. Only one out of the three sources was undetected in X-ray, while one had a $2\sigma$ detection, and the third one was not observed in the X-ray.
\citet{lin26} selected their sample from the SDSS DR17 spectroscopic database based on their 'v-shape' (i.e., they include LRDs only). They removed sources with strong \Nii\ and \Mgii\ emission, as possible type-I AGN contaminants, and required an \Oiii\ equivalent width above $ 10 \AA$ to reduce the contamination from galaxies. The objects are compact on a scale of 1.8\arcsec, and their final sample includes three objects at $z\sim 1$ selected as the most representative LRDs. Only one out of the three sources is undetected in X-ray, while one has a $2\sigma$ detection, and the third one was not observed in the X-ray.
\citet{park26} considered a sample of eight LRDs at $0.2 < z < 0.5$ that were selected from the DESI DR1 spectroscopic catalog. The sample includes both LRDs and LBDs and only two out of eight sources have an \Ha\ $FWHM > 1000$ \kms\ (i.e., our width cut). The sources are compact on a scale of 1\arcsec\, and their X-ray properties are not considered in the selection. 

% INCOMPLETENESS
More importantly, the samples of \citet{lin26} and \citet{park26} are incomplete. As noted by \citet{lin26}, the SDSS LRDs were disadvantaged in being selected for spectroscopic follow-ups due to their compact morphology. In addition, the adopted cut in \Oiii\ equivalent width may prevent some LRDs from being selected and that are similar to high-redshift LRDs reported in the literature (e.g., \citealt{ji25, naidu25, degraaff25}). According to \citet{park26}, it is very difficult to quantify the incompleteness of their sample, as their LRDs span a variety of selections and show up to the faintest magnitudes in the DESI catalog.
 
As discussed in \citet{lin26} and \citet{park26}, the measurements in the local universe should be regarded as lower limits to the number density of LRDs.
Future studies, based on blind spectroscopic samples of LRDs at $z \sim 0$, will be crucial to improve these estimates.

% POSSIBLE EXPLANATION, IF THE RESULT IS CONFIRMED
If the number density of LRDs is confirmed to decline sharply from cosmic noon to the present epoch, this may still be interpreted in the 
model framework proposed by \citet{inayoshi25}.  
%A possible explanation for an eventual decline of LRDs between cosmic noon and the present epoch could be related to the fact that AGNs act as LRDs in a limited phase of their life cycle.
%LRDs are likely a phase of the AGN life cycle. 
Indeed, assuming that LRDs represent the first one or two super-Eddington accretion episodes of newly-born black hole seeds, high accretion rates may become unsustainable once the black hole has grown considerably in mass and the gas supply from its host has decreased, possibly because of both the progressive consumption of the gas reservoir and AGN feedback. 
Therefore, objects that were shining as LRDs at cosmic noon would be observed as normal AGNs in the local universe.

\section{Summary and conclusions}
\label{sec:concl}

In this work, we report the quest for LRDs in the \textit{J1030 field} using NIRCam data (both imaging and spectroscopy) of the EIGER program. The main results of our study are summarized below:

\begin{itemize}

%LRDs in J1030
\item We discovered five point-like, broad-line emitters, undetected in X-rays. The sources have red color ($F200W-F356W > 0$) and are radio silent. Three of them are located at $z \sim 2.4$ (BiRD; \citealt{loiacono25}, ID 1017, and ID 3646) and were identified based on their \Hei\ and \Pag\ emission. The remaining two (ID 8511 and ID 9008) are \Ha\ emitters at $z\sim 4.5$. The $FWHM$ of the broad lines is in the range $\sim 1100 - 1700$ \kms , depending on the source, classifying them as BLAGNs.\\ 

% Black hole masses
\item Based on the broad \Ha\ line, we estimate the black hole masses of the sources. For the $z\sim 2.4$ emitters, the extinction-corrected \Ha\ luminosity was inferred from the \Pag\ luminosity, assuming case B recombination (see also \citealt{loiacono25}). We find black hole masses in the range of $10^7 - 10^8$ \Msun . The \Ha\ luminosities were also converted to the bolometric ones, which are $L_{\rm bol} \sim (3-33) \times 10^{44}\ \ergs$.\\  

% X-ray properties
\item We derived upper limits to the X-ray luminosity of the sources, which are in the range of $L_{\rm X} < (1.4 - 7.0) \times 10^{42}\ \ergs$. The stacked, median X-ray luminosities at $z \sim 2.4$ and $z\sim 4.5$ are $L_{\rm X} < 3.0 \times 10^{42}\ \ergs$ and $L_{\rm X} < 1.5 \times 10^{43}\ \ergs$, respectively. These values correspond to X-ray bolometric corrections $L_{\rm bol}/L_{\rm 2-10\ keV,rest} > 234$ and $L_{\rm bol}/L_{\rm 2-10\ keV,rest} > 25$, which are a factor $\sim 14$ and $\sim 2$ larger than the classical AGN median value.\\

% LRDs and LBDs
\item According to their optical and UV slopes, the BLAGNs can be split into three LRDs (BiRD, ID 8511, ID 9008), one LBD (ID 3646), and one intermediate object (ID 1017). As noted in other works, the prevalence of LRDs in our, admittedly small, sample could be due to the increasing fraction of LRDs towards higher \Ha\ luminosities, such as those sampled by our observations.\\

% LFs at z = 2.4 and z = 4.5
\item Without splitting the sample into sub-populations, we derived the bolometric LFs of LRDs at $z \sim 2.4$ and $z \sim 4.5$.
At cosmic noon, the space density of LRDs is a factor of $\sim 2-2.4$ lower than that of all classical AGNs with comparable bolometric luminosity \citep{shen20}, and consistent with the extrapolated abundance of UV-selected quasars \citep{kulkarni19}. 
At $z \sim 4.5$, the space density of LRDs is fully consistent with that estimated by other works for the same population \citep{matthee24, kokorev24}, and with that of pre-JWST AGNs \citep{shen20}.\\

% BHMFs at z = 2.4 and z = 4.5
\item We derive the abundance of LRDs as a function of the black hole mass. At $z\sim 2.4$, our estimate is consistent with that of BLQSOs inferred by \citet{shenkelly12}. At $z\sim 4.5$ our point is consistent with the result of \citet{matthee24}, although both estimates possibly suffer from incompleteness due to the $FWHM$ cut considered in both selections.\\

% Number density of LRDs at cosmic noon
\item Finally, we estimate the number density of LRDs at cosmic noon. The abundance of LRDs with $L_{\rm bol} > 3 \times 10^{44}\ \ergs$ at $z \sim 2.4$ is $n = 3.4^{+5.6}_{-2.4} \times 10^{-5}\ \rm Mpc^{-3}$. This implies that there is no significant evolution in the number density of LRDs with equal bolometric luminosity from $z \sim 7$ to $z \sim 2.4$. At cosmic noon, the number density of LRDs is a factor of $\sim 350$ larger than model predictions \citep{inayoshi25}. We also speculate that the $\sim 10\times$ higher density we find at $z \sim 2.4$ compared to studies based on ground-based data \citep{ma25} arises because those studies are likely sampling only the bright end of the LRD luminosity function.\\

% Comparison with other AGN populations
\item At cosmic noon, LRDs are a factor of $\sim 3.6$ less abundant than classical AGNs \citep{shen20}, whereas their number density equals that of X-ray (Compton-thin) selected AGNs with the same bolometric luminosities. While it is tempting to conclude that LRDs may be part of the missing population of Compton-thick AGNs, yet, because of the uncertainties in the interpretation of their X-ray weakness, it is difficult to quantify to what extent. 
\end{itemize}

Our results may imply that, if LRDs represent the early, rapid stages of growth of black holes \citep{inayoshi25}, the formation of (heavy) black hole seeds can be efficient even at epochs as recent as cosmic noon. Alternatively, our result may suggest that LRDs are likely a high-accretion phase of already mature black holes rather than the very initial accreting episodes of black hole seeds. Further studies will be necessary to distinguish between these two scenarios. 

\begin{acknowledgements} 
We thank Pierluigi Rinaldi for useful discussion. We acknowledge support from the INAF 2022/2023 "Ricerca Fondamentale" grants. KI acknowledges support under the grant PID2022-136828NB-C44 provided by MCIN/AEI/10.13039/501100011033 / FEDER, UE. I.J. and R.M. acknowledge support by the Science and Technology Facilities Council (STFC), by the ERC through Advanced Grant 695671 “QUENCH”, and by the UKRI Frontier Research grant RISEandFALL.
I.J. acknowledges support also by the Huo Family Foundation through a P.C. Ho PhD Studentship. 
R.M. also acknowledges funding
from a research professorship from the Royal Society. M.S. acknowledges financial support from the Italian Ministry for University and Research, through the grant PNRR-M4C2- I1.1-PRIN 2022-PE9-SEAWIND: Super-Eddington Accretion: Wind, INflow and Disk-F53D23001250006-NextGenerationEU. This work is based on observations made with the NASA/ESA/CSA James Webb Space Telescope. The data were obtained from the Mikulski Archive for Space Telescopes at the Space Telescope Science Institute, which is operated by the Association of Universities for Research in Astronomy, Inc., under NASA contract NAS 5-03127 for JWST. These observations are associated with the GTO program 1243.
\end{acknowledgements}

\bibliographystyle{aa}
\bibliography{main}

\begin{appendix}
\section{SEDs of the X-ray silent BLAGNs}
\label{app:sed}
We report the SEDs and the NIRCam composite images ($F115W + F200W + F356W$) of the compact, X-ray silent BLAGNs studied in this work. The SED and cutouts of BiRD are shown in \citet{loiacono25}.
The photometry includes ground-based measurements, JWST, and, when available, HST data-points ($F775W$, $F850LP$, and $F160W$). The $3\sigma$ upper limits are also shown.
For the ground-based data points of ID 3646, we used aperture-corrected magnitudes to derive the corresponding flux densities. This approach was necessary because contamination from a stellar spike makes the total magnitudes in the Y, J, and K bands reported in the J1030 catalog \citep{mazzolari26} unreliable. Since the source is point-like, aperture-corrected magnitudes are perfectly suited to derive the total fluxes of the target.
\begin{figure*}
\begin{center}
\includegraphics[width=1\textwidth]{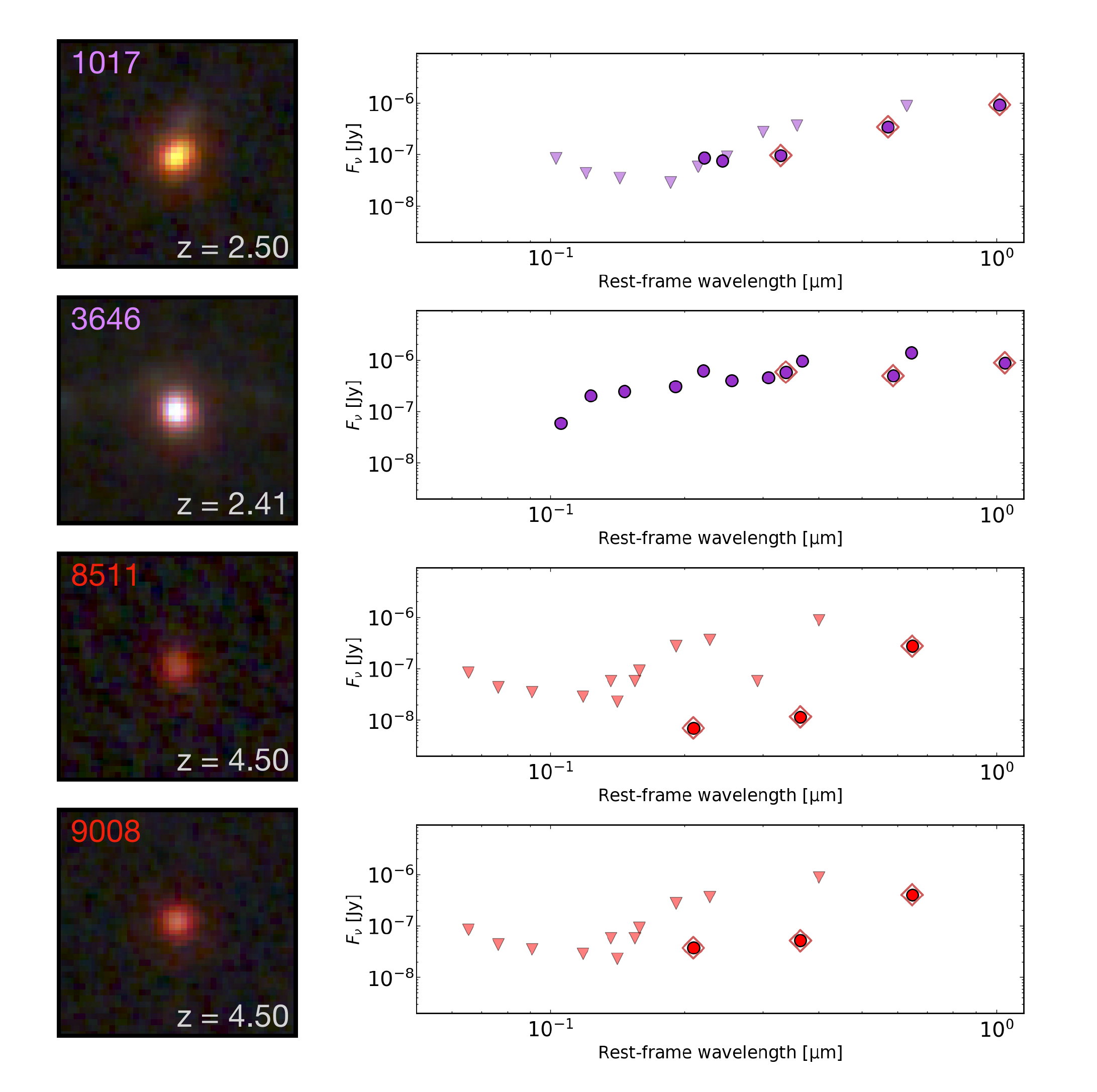}\\
\end{center}
\caption{NIRCam 1.2\arcsec $\times$ 1.2\arcsec\ cutout and SED of ID 1017, ID 3646, ID 8511 and ID 9008. The cutouts were obtained adding the $F115W + F200W + F356W$ images. The red diamonds enclose the JWST flux densities. The inverted triangles correspond to the $3\sigma$ upper limits.}
\label{fig:sed}
\end{figure*}
\end{appendix}
\end{document}